%% file: main.tex
\documentclass[sigconf]{acmart}
\usepackage{tabularx} 
\usepackage{multirow} 
\usepackage{bm} 
\usepackage{tikz} 
\usepackage{tabularx} 
\usepackage{xspace} 
\usepackage{makecell} 
\usepackage{adjustbox} 
\usepackage[nameinlink,capitalize]{cleveref}
\usetikzlibrary{decorations.pathreplacing, arrows,positioning, shapes.misc, backgrounds}
\usepackage{xparse}

\NewDocumentCommand{\scnum}{ >{\SplitArgument{1}{e}}m }
 {\scnumaux#1}
\NewDocumentCommand{\scnumaux}{ m m }
 {#1\,\mathrm{e}{#2}}

\newcolumntype{R}[2]{%
    >{\adjustbox{angle=#1,lap=\width-(#2)}\bgroup}%
    l%
    <{\egroup}%
}

\AtBeginDocument{%
  \providecommand\BibTeX{{%
    \normalfont B\kern-0.5em{\scshape i\kern-0.25em b}\kern-0.8em\TeX}}}

\setcopyright{acmcopyright}

\copyrightyear{2021}
\acmYear{2021}
\setcopyright{iw3c2w3}
\acmConference[WWW '21]{Proceedings of the Web Conference 2021}{April 19--23, 2021}{Ljubljana, Slovenia}
\acmBooktitle{Proceedings of the Web Conference 2021 (WWW '21), April 19--23, 2021, Ljubljana, Slovenia}
\acmPrice{}
\acmDOI{10.1145/3442381.3449945}
\acmISBN{978-1-4503-8312-7/21/04}

\begin{document}
\newcommand{\modelname}{{Radflow}\xspace}

\newcommand{\modelmean}{\modelname-MeanPooling\xspace}
\newcommand{\modelsage}{\modelname-GraphSage\xspace}
\newcommand{\modelgat}{\modelname-GAT\xspace}

\newcommand{\Vevo}{\textsc{VevoMusic}\xspace}
\newcommand{\Wiki}{\textsc{WikiTraffic}\xspace}
\newcommand{\LA}{\textsc{Los-loop}\xspace}
\newcommand{\SZ}{\textsc{SZ-taxi}\xspace}

\newcommand{\bx}{\bm{x}}
\newcommand{\bX}{\bm{X}}
\newcommand{\bA}{\bm{A}}
\newcommand{\by}{\bm{y}}
\newcommand{\bv}{\bm{v}}
\newcommand{\bV}{\bm{V}}
\newcommand{\bE}{\bm{E}}
\newcommand{\hv}{\hat{v}}
\newcommand{\hbv}{\hat{\bm{v}}}
\newcommand{\bw}{\bm{w}}
\newcommand{\tbw}{\tilde{\bm{w}}}
\newcommand{\bb}{\bm{b}}
\newcommand{\bz}{\bm{z}}
\newcommand{\bff}{\bm{f}}
\newcommand{\hbf}{\hat{\bm{f}}}
\newcommand{\hf}{\hat{f}}
\newcommand{\bh}{\bm{h}}
\newcommand{\bp}{\bm{p}}
\newcommand{\bi}{\bm{i}}
\newcommand{\bo}{\bm{o}}
\newcommand{\bu}{\bm{u}}
\newcommand{\hbu}{\hat{\bm{u}}}
\newcommand{\tbu}{\tilde{\bm{u}}}
\newcommand{\hbp}{\hat{\bm{p}}}
\newcommand{\bq}{\bm{q}}
\newcommand{\hbq}{\hat{\bm{q}}}
\newcommand{\bc}{\bm{c}}
\newcommand{\tbc}{\tilde{\bm{c}}}
\newcommand{\bW}{\bm{W}}
\newcommand{\bU}{\bm{U}}
\newcommand{\LSTM}{\text{LSTMCell}}
\newcommand{\Block}{\text{Block}}
\newcommand{\FF}{\text{FeedForward}}
\newcommand{\FM}{\text{ForecastModel}}
\newcommand{\st}{~\text{s.t.}~}
\newcommand{\GELU}{\text{GELU}}
\newcommand{\softmax}{\text{Softmax}}
\newcommand{\R}{\mathbb{R}}
\newcommand{\N}{\mathcal{N}}
\newcommand{\cV}{\mathcal{V}}
\newcommand{\T}{\mathcal{T}}

\newcommand{\eat}[1]{}
\newcommand{\rev}[1]{{\color{blue}{#1}}}
\newcommand{\rvx}[1]{{\color{red}{#1}}}
\newcommand{\alasdair}[1]{{\color{blue}{#1}}}
\newcommand{\verify}[1]{{\bf\color{red}{#1}}}



\title{\modelname: Decomposable Time Series Modeling with Network Information}
\title[\modelname]{\modelname: A Recurrent, Aggregated, and Decomposable Model for Networks of Time Series \vspace{-0.5em}}



\author{Alasdair Tran$^{1,2}$, Alexander Mathews$^{1}$, Cheng Soon Ong$^{1,2}$, Lexing Xie$^{1}$}
\affiliation{%
\institution{$^1$ Australian National University $\qquad$ $^2$ Data61, CSIRO}
}
\email{{alasdair.tran,alex.mathews,chengsoon.ong,lexing.xie}@anu.edu.au}

\renewcommand{\shortauthors}{Alasdair Tran, Alexander Mathews, Cheng Soon Ong, Lexing Xie}

\input{01_abstract}
\maketitle

\input{02_intro}
\input{03_related_works}
\input{05_problem}
\input{04_model_diagram}
\input{06_model}

\input{07_dataset}
\input{08_experiments}
\input{09_results}
\input{10_conclusion}

\begin{acks}
    This research is supported in part by the Australian Research Council
    Project DP180101985 and AOARD project 20IOA064. We thank NVIDIA for
    providing us with Titan V GPUs for experimentation.
\end{acks}

\bibliographystyle{ACM-Reference-Format}
\bibliography{main}

\clearpage
\input{appendix}

\end{document}

%% file: 01_abstract.tex


\begin{abstract}
We propose a new model for networks of time series that influence each other.
Graph structures among time series are found in diverse domains, such as web
traffic influenced by hyperlinks, product sales influenced by recommendation,
or urban transport volume influenced by road networks and weather. There has
been recent progress in graph modeling and in time series forecasting,
respectively, but an expressive and scalable approach for a network of series
does not yet exist. We introduce \modelname, a novel model that embodies three
key ideas:
a recurrent neural network to obtain node embeddings that depend on time, the
aggregation of the flow of influence from neighboring nodes with multi-head
attention, and the multi-layer decomposition of time series. \modelname
naturally takes into account dynamic networks where nodes and edges change over
time, and it can be used for prediction and data imputation tasks. On
real-world datasets ranging from a few hundred to a few hundred thousand nodes,
we observe that \modelname variants are the best performing model across a wide
range of settings. The recurrent component in \modelname also outperforms
N-BEATS, the state-of-the-art time series model. We show that \modelname can
learn different trends and seasonal patterns, that it is robust to missing
nodes and edges, and that correlated temporal patterns among network neighbors
reflect influence strength. We curate \Wiki, the largest dynamic network of
time series with 366K nodes and 22M time-dependent links spanning five years.
This dataset provides an open benchmark for developing models in this area,
with applications that include optimizing resources for the web. More broadly,
\modelname has the potential to improve forecasts in correlated time series
networks such as the stock market, and impute missing measurements in
geographically dispersed networks of natural phenomena.

{\bgroup
  \par\medskip\small\noindent{\bfseries ACM Reference Format:}\par\nobreak
  \noindent\bgroup
  Alasdair Tran, Alexander Mathews, Cheng Soon Ong, Lexing Xie. 2021.
  \modelname: A Recurrent, Aggregated, and Decomposable Model for Networks of Time
  Series. In {\itshape Proceedings of the Web Conference 2021 (WWW '21), April
  19–23, 2021, Ljubljana, Slovenia}. ACM, New York, NY, USA, 12 pages.
  https://doi.org/10.1145/3442381.3449945
\par\egroup}
\end{abstract}

\keywords{time series, networks, graphs, sequence models, wikipedia}

%% file: 02_intro.tex
\section{Introduction}

\begin{figure*}[!t]
    \includegraphics[width=\textwidth]{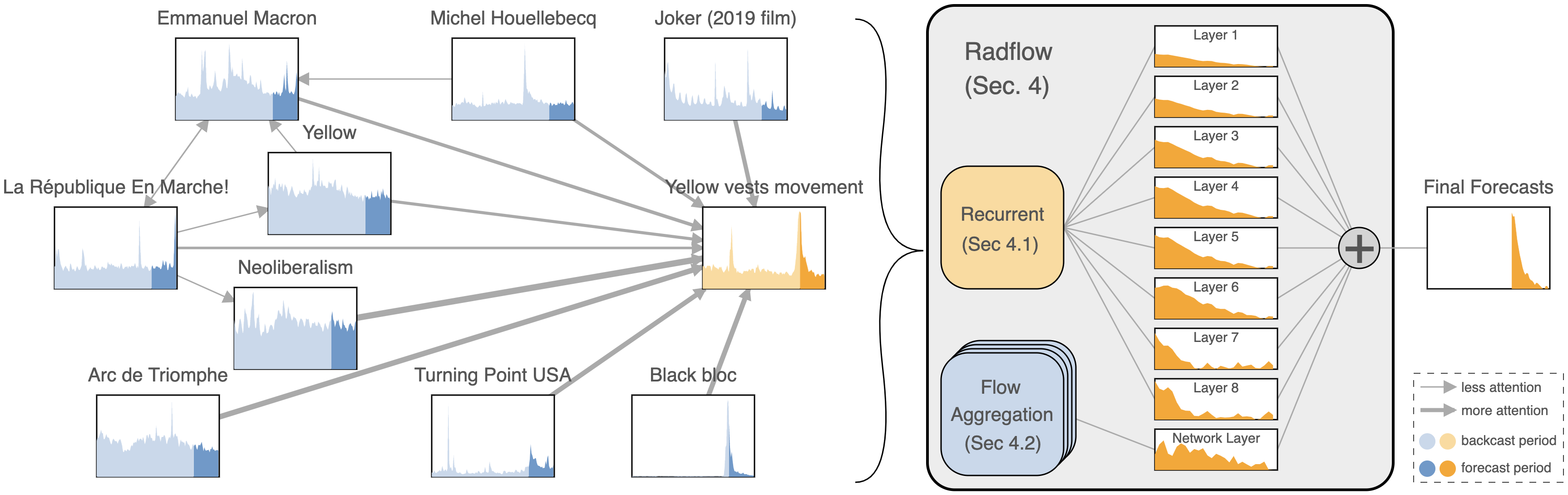}
    \caption{Overview of the \modelname model, centered around the \Wiki
     subgraph of
     \href{https://en.wikipedia.org/wiki/Yellow_vests_movement}{\textit{Yellow
     vests movement}} (a social movement in France since 2018, shown on the
     left). Each node is a page with a time series of view counts, shown in its
     individual mini-panel. Edge strengths correspond to average attention
     scores in the forecast period. The final forecast is produced by summing
     over the output from eight layers of the recurrent component and the
     network flow aggregation component (shown on the right). \modelname
     correctly predicts a sharp drop in traffic volume for \textit{Yellow vests
     movement} from 3 June 2020 to 30 June 2020, due to information from
     neighboring nodes such as \textit{Black bloc} and \textit{Turning Point
     USA}.}

    \Description{A subgraph centered around the Wikipedia page ``Yellow vest
    movement''.}
    \label{fig:teaser}

\end{figure*}

Forecasting time series is a long-standing research problem that is applicable
to econometrics, marketing, astronomy, ocean sciences, and other domains.
Similarly, networks
are the subject of active research with broad relevance to areas such as
transportation, internet infrastructure, signaling in biological processes, and
online media. In this paper, we are concerned with forecasting among a network
of time series with mutual influences. Tools for tackling this problem will
help answer questions about complex systems evolving over time in the above
application domains and beyond.

In the context of recent progress in end-to-end learning for time series and
large networks, there are three prominent challenges. The first is {\em
expressiveness}, i.e. building models to represent richer classes of functions.
Recent works in predicting book sales~\cite{dhar2014prediction} and online
video views~\cite{Wu2019EstimatingAF} employed simple
aggregation~\cite{dhar2014prediction} or linear
combinations~\cite{Wu2019EstimatingAF} of the last observation of incoming
nodes. Recent graph neural
networks~\cite{Hamilton2017GraphSAGE,Velickovic2018GAT,Kipf2017GCN} provide
flexible aggregation among network neighbors, but do not readily apply to time
series. N-BEATS~\citep{Oreshkin2020NBEATS} is the new state-of-the-art model in
time series benchmarks, using a stack of neural modules to decompose history;
but this architecture does not provide a usable representation for a network of
series as the neural modules do not explicitly encode the temporal structure of
the data. There are several graph-to-sequence
tasks~\cite{Xu2018Graph2seq,Cai2020GraphTransformer} considered in the natural language
processing (NLP) domain, but the networks of time series problem, cast in such
terminology, is graphs-of-sequences to graphs-of-sequences.

The second challenge is {\em scale}. Our goal is to model longitudinal (e.g.,
daily) time series spanning a few years, and large networks in the order of
hundreds of thousands of nodes. This requires scalability in the time series
component, the graph component, as well as their interactions. The recently
proposed T-GCN model~\cite{Zhao2019TGCNAT}, for example, nests a graph neural
network within a recurrent neural network, which is limited in both space and
time complexity that prevents it from scaling to web-scale networks. The
networks used in their evaluation contain only a few hundred nodes.

The third is the {\em dynamic nature of links and nodes in the network}. For
example, \citet{Wu2019EstimatingAF} reported that 50\% of online video
recommendation links appear in fewer than 5 out of 63 days of observation, and
we observe that more than 100K new Wikipedia pages were created over the first
half of 2020 alone. Dynamic networks are an active topic of attention for graph
neural
networks~\citep{Manessi2020DynamicGC,Pareja2019EvolveGCNEG,trivedi2019dyrep},
but these existing algorithms are designed for link prediction and node
classification, not for time series forecasting.


We propose a novel neural model for networks of time series that tackles
all three challenges.
We adopt a {\bf R}ecurrent structure that affords time-sensitive {\bf
A}ggregations of network flow on top of the {\bf D}ecomposition principle of
time series; hence the name {\bf\modelname}. It is more expressive than
N-BEATS~\cite{Oreshkin2020NBEATS} because it can generate node embeddings to
handle graph inputs. It is more scalable than T-GCN as it can process hundreds
of thousands of nodes via network attention and importance-weighted node
sampling.
The structure of \modelname allows it to take dynamically changing nodes and
edges as inputs, makes it tolerant to missing data, and is suitable for
multivariate time series.
Moreover, its multi-head attention strategy and layered decompositions provide
interpretations over network influences and time.

\modelname is evaluated on four datasets. Two are on urban traffic consisting
of several hundred nodes; two are large-scale datasets---\Vevo containing 61K
videos~\cite{Wu2019EstimatingAF} and a newly curated \Wiki dataset containing
366K pages and 22M dynamic links. On both \Vevo and \Wiki, \modelname without
network information is consistently better than the comparable
N-BEATS~\cite{Oreshkin2020NBEATS}. Among models with network information,
\modelname variants perform the best in both imputation and forecasting tasks.
In particular, \modelname outperforms state-of-the-art
ARNet~\cite{Wu2019EstimatingAF} by 19\% in SMAPE score on \Vevo. We find that
the layers in the recurrent component capture different seasonality and trends,
while attention over the network captures the time-varying influence from
neighboring nodes. \cref{fig:teaser} illustrates the task of predicting 28 days
of view counts on {\it Yellow vests movement}, based on the historical traffic
of that page and the traffic of the neighboring pages. \modelname correctly
predicts the sharp drop that is observed during the test period.

Our key contributions include:
\begin{enumerate}
    \vspace{-0.2em}
    \item \modelname, an end-to-end neural forecasting model for dynamic
    networks of multivariate time series that is scalable to hundreds of
    thousands of nodes. 

    \item Interpretable predictions over time series patterns via layered
    decompositions, and over network neighbors via multi-head attention.


    \item Consistently outperforming state-of-the-art time series forecasting
    models and networked series forecasting models in real-world datasets
    across a diverse set of tasks.

    \item \Wiki, the largest dynamic network of time series, containing
    multi-dimensional traffic data from 366K Wikipedia pages and 22M dynamic
    links over five years. The dataset, code, and pretrained models are
    available on
    GitHub\footnote{\href{https://github.com/alasdairtran/radflow}{https://github.com/alasdairtran/radflow}}.
\end{enumerate}

%% file: 03_related_works.tex
\section{Related Works}

\subsection{Time series modeling}

Time series modeling has an extensive literature spanning many fields.
Classical approaches~\cite{hyndam2018forecast} include exponential smoothing
and autoregressive integrated moving average (ARIMA) models. Exponential
smoothing uses exponentially decaying weights of past observations, while
extensions can incorporate both trends and
seasonality~\cite{holt1957forecasting, brown1959statistical,
winters1960forecasting}. ARIMA~\cite{box1970arima} models aim to describe
auto-correlations in a time series using a linear combination of past
observations and forecasting errors; extensions can also incorporate
seasonality. In recent years, neural network approaches have become more
popular. \citet{Wu2020DeepTM} used transformers to forecast influenza
activities. \citet{Zhu2017DeepAC} used a Bayesian neural network to model
uncertainty in the forecasting.

\citet{Oreshkin2020NBEATS} were the first to show that a pure neural model
without any time series-specific component can outperform existing statistical
techniques on the benchmark datasets \textsc{M3}~\cite{Makridakis2000M3},
\textsc{M4}~\cite{Makridakis2018M4}, and
\textsc{Tourism}~\cite{Athanasopoulos2011Tourism}. Their proposed
model, N-BEATS, treats the time series prediction as a non-linear multivariate
regression problem that outputs a fixed-length vector as the forecast. N-BEATS'
key modeling component is the stacking of layers, each of which takes as input
the residual time series calculated by the previous layers.
However, N-BEATS only works with one-dimensional time series and does not
produce a time series representation vector at each step, making it difficult
to be used in settings with dynamic network information.
We address both of these shortcomings by adopting the recurrent network
structure that produces time series representations (which we call embeddings)
at any time step, while still taking advantage of the residual stacking idea
from N-BEATS.

Another type of time series is discrete events happening in continuous time,
often described by temporal point processes. Predictive approaches using point
processes~\cite{mishra2016feature} require the data to contain details of
individual events rather than the aggregate statistics that are more common in
large-scale web data due to privacy and storage constraints. Point process
estimation models are typically quadratic in the number of events, which is too
expensive for large-scale data like \Vevo and \Wiki.

\subsection{Prediction over networks}

Network effects in online services are an active area of research that studies
how links between online items determine properties such as visibility,
influence, and future behavior. Social networks and information networks are
frequently studied in this context. On Twitter, \citet{su2016effect} showed
that the introduction of a new network-based recommender system resulted in a
substantial change to the network structure, exacerbating the ``rich get
richer'' phenomenon. On Wikipedia, the link structure has been used to track the
evolution of emerging topics~\cite{kampf2015detection} and the flow of traffic
caused by exogenous events~\cite{zhu2020content}. \citet{zhu2020content} showed
that when a Wikipedia article gains attention from an exogenous event, it can
lead to a substantial rise in attention on downstream hyperlinked articles.
\citet{kampf2015detection} showed that the evolution of an emerging topic
can be tracked and predicted using Wikipedia page views and
internal links. The product recommendation network on Amazon has been shown to
affect purchasing decisions~\cite{oestreicher2013network}.
\citet{Wu2019EstimatingAF} showed that the network induced by YouTube's
recommender system leads to a flow of influence between music videos.

One recent approach to model the network effect is to use neural graph networks
to generate low-dimensional embeddings of nodes in a graph. Early methods such
as node2vec~\cite{Grover2016node2vec} and DeepWalk~\cite{Perozzi2014DeepWalk}
are transductive, designed mainly to work with a fixed graph. Recent models can
be applied to an inductive setting that requires generating embeddings for
nodes not seen during training. This is done, for example, by sampling and
aggregating node features from the local neighborhood in
GraphSage~\cite{Hamilton2017GraphSAGE}. Various aggregation methods have been
proposed including max-pooling~\cite{Hamilton2017GraphSAGE} and
mean-pooling~\cite{Kipf2017GCN}. \citet{Velickovic2018GAT} proposed Graph
Attention Network (GAT) that uses a modified version of the multi-head
attention~\cite{Vaswani2017AttentionIA} to aggregate the neighborhood. In our
proposed architecture, network embedding and aggregation are key components. A
variety of network aggregation mechanisms can be used (see \cref{ssec:flow} and
\cref{sec:evalsettings}). Our suggested aggregation mechanism is similar to
GAT, but instead uses the original, and more common, dot-product formulation
of the multi-head attention.

\subsection{Networks of time series}

Being a new area of research, there are few forecasting methods and a limited
number of datasets for networks of time series. Early approaches ignore the
network structure and instead treat each node as an independent
series~\cite{szabo2010predicting,Petluri2018WebTraffic}.
\citet{Wu2019EstimatingAF} incorporated the local network structure into an
autoregressive time series model, but the architecture only works with a
static graph.
\citet{Zhao2019TGCNAT}
proposed a new Recurrent Neural Network (RNN) cell called T-GCN that takes into
account the structure of a static graph by incorporating a Graph Convolution
Network (GCN) component. The bundling of these two components and the lack of
neighborhood sampling makes T-GCN too computationally expensive to apply to
graphs of more than a few hundred nodes.

A related problem is predicting how edges in networks change, such as those
using point processes~\cite{trivedi2017know,trivedi2019dyrep} or
two-dimensional attention over graphs and time~\cite{sankar2020dysat}. We do
not tackle this problem; instead we assume that the dynamic graph is observed
(such as generated by a recommender system or crowd-sourcing), and the
prediction target is a time series on each node rather than the evolving graph
itself. Our work is the first forecasting model optimized for large dynamic
networks of time series.



%% file: 05_problem.tex
\section{Problem Statement}
\label{sec:problem}
Consider the problem of time series forecasting in a graph. The input is a
graph $G = (\bV, \bE)$, consisting of $N$ nodes denoted $\bV = \{\bv^1,
\bv^2,\dots,\bv^N \}$, and $M$ edges. Each node $\bv^j$ is associated with a
multivariate time series having $T$ observations:
\begin{align}
    \bv^j &= [\bv^j_{1}, \bv^j_{2}, \dots, \bv^j_{T}]
\end{align}
where $\bv^j_{t} \in \R^D$ is the $D$-dimensional observation vector of node
$\bv^j$ at time step $t$. When the time series has only one value per time step
(univariate), then $D=1$. We use $\bv^j_{[t:s]}$ to denote a subsequence of
$\bv^j$ containing all observations from time $t$ to $s$, where $t \leq s$:
\begin{align}
    \bv^j_{[t:s]} &= [\bv^j_{t}, \bv^j_{t+1}, \dots, \bv^j_{s-1}, \bv^j_{s}]
\end{align}
If node $\bv^{i}$ has the potential to directly influence the time series of
node $\bv^{j}$ at time step $t$, then we add a directed edge $e^{ij}_{t}$ from
$\bv^{i}$ to $\bv^{j}$, and $\bv^{i}$ becomes a neighbor of $\bv^{j}$. We
define $\N_t(\bv^{j})$ to be the set of neighbors of $\bv^{j}$ at times step
$t$. Edges may appear and disappear over time, thus $G$ is a dynamic graph. We
can now represent $G$ as an adjacency array $\bA \in \R^{N
\times N \times T}$.
For an unweighted directed graph, the entry $a_{ijt}$ in
$\bA$ is 1 if edge $e^{ij}_{t}$ exists, and 0 otherwise.

We now define the time series forecasting problem as it applies to dynamic
graphs. The forecast length $F$ is the number of future time steps for which
the model will make predictions, while the backcast length $B$ is the number of
past observations available for making such predictions. Suppose we are
currently at time $t=0$. To forecast the time series of node $\bv^j$
(which we shall call the \textit{ego} node)
from time
step $1$ to $F$, the forecast model
\begin{align}
    \hbv^j_{[1,F]}  = \FM \left(\bv^j_{[-B+1:0]}, \cV^{\N(\bv^j)}\right)
\end{align}
will take two sets of inputs: the $B$ most recent observations of $\bv^j$ and
the information from its neighbors. This leads to two different settings, both
of which will be evaluated in \cref{sec:results}. The first is
\textsc{Imputation}, in which we observe the true values of the neighbors at
the time of prediction. This amounts to using the ground-truth observations of
the neighbors during the forecast period:
\begin{align}
    \cV^{\N(\bv^j)} = \left\{ \bv'_{[-B+1:F]} \,\,\middle|\,\, \bv' \in \N(\bv^j) \right\}
\end{align}
This is the setting used by \citet{Wu2019EstimatingAF} and is most useful when
the main goal is to fill in missing data in the time series, or to interpret
the influence between nodes. The second setting is \textsc{Forecast}, where we
first use our best pure time series model to predict the future observations of
each neighbor. These predictions are then used in the full model to forecast
$\bv^j$ itself.

In both settings, the final output of the model is
\begin{align}
    \hbv^j_{[1,F]} = [\hbv^j_{1}, \hbv^j_{2}, \dots, \hbv^j_{F}]
\end{align}
corresponding to the forecast values for the next $F$ time steps. Here we use
the hat notation to denote the model's predictions, e.g., $\hbv^j_{t}$ is the
forecast time series vector of the ground truth $\bv^j_t$.

%% file: 04_model_diagram.tex
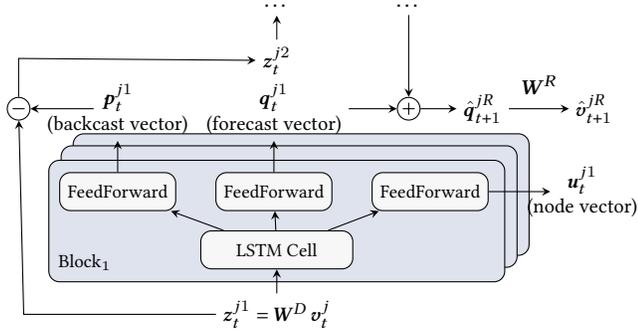
\begin{figure}[t]
    \centering
    \definecolor{cyan}{RGB}{223,227,238}
    \definecolor{orange}{RGB}{255,223,186}
    \definecolor{white}{RGB}{247,247,247}
    \tikzstyle{arrow} = [->,>=stealth]
    \tikzstyle{eqn} = [scale=0.85]
    \tikzstyle{blockrect} = [rounded corners, minimum width=6cm, minimum height=1.6cm, text centered, draw=black, fill=cyan]
    \tikzstyle{lstmrect} = [rounded corners, minimum width=2cm, minimum height=0.5cm, text centered, draw=black, fill=white]
    \tikzstyle{ffrect} = [rounded corners, minimum width=1cm, minimum height=0.5cm, text centered, draw=black, fill=white]
    \tikzset{%
        do path picture/.style={%
            path picture={%
            \pgfpointdiff{\pgfpointanchor{path picture bounding box}{south west}}%
                {\pgfpointanchor{path picture bounding box}{north east}}%
            \pgfgetlastxy\x\y%
            \tikzset{x=\x/2,y=\y/2}%
            #1
            }
        },
        plus/.style={do path picture={
            \draw [line cap=round] (-1/2,0) -- (1/2,0) (0,-1/2) -- (0,1/2);
        }},
        minus/.style={do path picture={
            \draw [line cap=round] (-1/2,0) -- (1/2,0) ;
        }}
    }
    \resizebox{\columnwidth}{!}{%
    \begin{tikzpicture}

        \node(Block1) [blockrect] {};
        \begin{scope}[on background layer]
        \node(Block2) [blockrect,above right=-4em and -18em of Block1, node distance=1.2em] {};
        \node(Block2) [blockrect,above right=-4.5em and -18.5em of Block1, node distance=1.2em] {};
        \end{scope}
        \node(LSTM1) [lstmrect, below of=Block1, node distance=1.2em] {\footnotesize LSTM Cell};

        \node(FF1) [ffrect, above left=0.8 em and 1em of LSTM1] {\footnotesize FeedForward};
        \node(FF2) [ffrect, above left=0.8 em and -5.5em of LSTM1] {\footnotesize FeedForward};
        \node(FF3) [ffrect, above left=0.8 em and -12em of LSTM1] {\footnotesize FeedForward};

        \draw[arrow] (LSTM1) -- (FF1);
        \draw[arrow] (LSTM1) -- (FF2);
        \draw[arrow] (LSTM1) -- (FF3);

        \node[eqn,above=1.3em of FF1, align=center] (p1) {$\bp_t^{j1}$ \\(backcast vector)};
        \draw[arrow] (FF1) -- (p1);

        \node[eqn, above=1.3em of FF2, align=center] (q1) {$\bq_t^{j1}$ \\(forecast vector)};
        \draw[arrow] (FF2) -- (q1);

        \node[eqn, right=1.3em of FF3, align=center] (u1) {$\bu_t^{j1}$ \\(node vector)};
        \draw[arrow] (FF3) --++(5em,0) (u1);

        \node[eqn, below=1em of LSTM1] (z1) {$\bz^{j1}_{t} = \bW^D \, \bv^j_{t}$};
        \draw[arrow] (z1) -- (LSTM1);

        \node[eqn, above=6.2em of LSTM1] (z2) {$\bz^{j2}_{t}$};

        \node [circle, draw, minus, left=0.4em of p1] (minus1) {};
        \coordinate[above = 1.6em of minus1] (corner1);
        \draw[arrow] (p1) ++(-1em,0)-- (minus1);
        \draw[arrow] (minus1) -- (corner1) -- (z2);

        \coordinate[left = 8em of z1] (corner2);
        \draw[arrow] (z1) -- (corner2) -- (minus1);

        \node (dots1) [above=1em of z2] {\dots};
        \draw[arrow] (z2) -- (dots1);

        \node [circle, draw, plus, right= 2em of q1] (plus1) {};
        \node[eqn, right=1.5em of plus1] (qh1) {$\hbq^{jR}_{t+1}$};
        \draw[arrow] (q1) -- (plus1);
        \draw[arrow] (plus1) -- (qh1);

        \node (dots2) [above=3.41em of plus1] {\dots};
        \draw[arrow] (dots2) -- (plus1);

        \node[eqn, right=2.5em of qh1] (vh1) {$\hbv^{jR}_{t+1}$};
        \draw [eqn, arrow] (qh1) -- node[name=wr,yshift=1em] {\small $\bW^R$} (vh1);

        \node(label1) [below left=-1.5em and -3em of Block1] {\footnotesize $\Block_1$};

    \end{tikzpicture}}%

    \caption{An overview of \modelname's recurrent block
    (\cref{ssec:recurrent}). Shown here is the first of L blocks at time $t$,
    which takes a projected representation $z_t^{j1}$ of the raw observations
    as input, and produces: the backcast vector $\bp_t^{j1}$,
    the forecast vector $\bq_t^{j1}$, and the node vector $\bu_t^{j1}$. The
    backcast vector is subtracted from $z_t^{j1}$ to obtain the (residual)
    input for the next block. The forecast vector is an additive component for
    the overall forecast $\hbv^{jR}_{t+1}$ in \cref{eq:vhat_from_qhat}. The
    node vector will be used when aggregating the neighbors
    (\cref{ssec:flow}).}

    \Description{An overview of the first recurrent block in \modelname.}
    \label{fig:recurrent_comp}
  \end{figure}

%% file: 06_model.tex

\section{\modelname}
\label{sec:radflow}

\modelname consists of two main modules: a recurrent component and a flow
aggregation component. The recurrent component models all the time series in
the graph independently, while the flow aggregation component additively
adjusts the predictions based on the neighboring time series. The forecast
$\hbv^{j}_{t}$ of node $\bv^j$ at time step $t$ is obtained by summing up the
outputs of the two main modules,
\begin{align}
    \hbv^{j}_{t} &= \hbv^{jR}_{t} + \hbv^{jA}_{t} \label{eq:sum}
\end{align}
where $\hbv^{jR}_{t}$ is the forecast contribution from the recurrent component
and $\hbv^{jA}_{t}$ is the contribution from the aggregation component. Note
that $\hbv^{jA}_{t}$ is itself a function of $\hbv^{jR}_{t}$.

\subsection{Recurrent component}
\label{ssec:recurrent}

We predict time series by breaking them down into $L$ components using stacked
recurrent blocks. The recurrent component is also designed to feed into the
flow aggregation component which uses the node vectors to aggregate information
in a neighborhood. \cref{fig:recurrent_comp} shows a
schematic diagram of the recurrent component.

We first project the historical observations of the time series into a latent
space in $\R^H$, where $H$ is the hidden state size:
\begin{align}
    \bz^{j1}_{t} &= \bW^D \, \bv^j_{t} \label{eq:input}
\end{align}
Here $\bW^D \in \R^{H \times D}$ is a learnable weight matrix. To get an
intuitive justification for this projection, consider the special case where
$D=1$ and $\bW^D$ is the all-ones vector. Then $\bz^{j1}_{t}$ would contain $H$
copies of the observation $v^j_{t}$. This resembles running an ensemble of $H$
different time series models in parallel.

The recurrent component of our model consists of $L$ blocks. Let $\bz^{j
\ell}_{t}$ be the input to $\Block_{\ell}$ for node $\bv^j$ at step $t$. In
particular, the vector $\bz^{j1}_{t}$ computed in \cref{eq:input} will be
used as the input to the first block. Each block will output three vectors---the
backcast vector $\bp^{j \ell}_{t}$, the forecast vector $\bq^{j \ell}_{t}$,
and the node vector $\bu^{j \ell}_{t}$:
\begin{align}
    (\bp^{j \ell}_{t},\, \bq^{j \ell}_{t}, \bu^{j \ell}_{t}) &= \Block_{\ell}(\bz^{j \ell}_{t})
    \label{eq:block}
\end{align}
with $\bp^{j \ell}_{t},\, \bq^{j \ell}_{t}, \bu^{j \ell}_{t} \in \R^H$.
Specifically inside each block, we have an LSTM cell followed by feedforward
layers. The LSTM cell is first to operate, accepting as input: the time series
residual computed by the previous block $\bz^{j \ell}_{t} \in \R^H$, the
previous time step's hidden state $\bh^{j \ell}_{t-1} \in \R^H$, and the cell
state $\bc^{j \ell}_{t-1} \in \R^H$. The LSTM cell produces a hidden output
$\bh^{j \ell}_{t}$, which is then passed through three different feedforward
layers:
\begin{align}
    \bp^{j \ell}_{t} &= \FF^{P \ell}(\bh^{j \ell}_{t}) \\
    \bq^{j \ell}_{t} &= \FF^{Q \ell}(\bh^{j \ell}_{t}) \\
    \bu^{j \ell}_{t} &= \FF^{U \ell}(\bh^{j \ell}_{t})
\end{align}
Each of the feedforward layers consists of two linear projections with a GELU
activation after the first linear projection:
\begin{align}
    \FF(\bh) &= \bW^{FF_2} \, \GELU( \bW^{FF_1} \, \bh)
\end{align}
The GELU activation function is a stochastic variant of ReLU that has been
shown to outperform ReLU in sequence-to-sequence
models~\cite{Hendrycks2016GaussianEL}. It is defined as
    $\GELU(x) = x \, \Phi(x) $,
where $\Phi$ is the standard Gaussian cumulative distribution function.

The
first output $\bp^{j \ell}_{t}$ is a component of the projected time series
captured by $\Block_{\ell}$. Subsequent blocks depend on the residual value of
the projected time series after removing this component:
\begin{align}
    \bz^{j \ell + 1}_{t} &=  \bz^{j \ell}_{t} - \bp^{j \ell}_{t}
\end{align}
The second output $\bq^{j \ell}_{t}$ is $\Block_{\ell}$'s contribution to the
forecast of the next time step. The final forecast representation of the
recurrent component will be the sum of all the blocks
\begin{align}
    \hbq^{jR}_{t+1} &= \sum_{\ell = 1}^{L} \bq^{j \ell}_{t}
    \label{eq:sum_of_q}
\end{align}
where $\hbq^{jR}_{t+1} \in \R^H$. We then project this into $\R^D$ to get the
forecast contribution from the recurrent component, i.e. the first term in
\cref{eq:sum}:
\begin{align}
    \hbv^{jR}_{t+1} &= \bW^R \, \hbq^{jR}_{t+1}
    \label{eq:vhat_from_qhat}
\end{align}

\subsection{Flow aggregation component}
\label{ssec:flow}

The flow aggregation component models the influence between the time series of
neighboring nodes in the network. This component takes as input time-dependent
embeddings from the recurrent component of each node in the neighborhood, and
produces as output the second term in \cref{eq:sum}. Each embedding summarizes
the time series of the corresponding node up to the current time. Let $\bu^j_t$
be the embedding of the ego node $\bv^j$ at time step $t$, formed by summing
the node vectors $\bu^{j \ell}_{t}$ over all $L$ blocks:
\begin{align}
    \bu^j_t = \sum_{\ell=1}^{L} \, \bu^{j \ell}_{t} \label{eq:node_embed}
\end{align}
In the \textsc{Imputation} setting, the set of embeddings of all neighbors of
the ego at time $t + 1$ is
\begin{align}
    \mathcal{U}^{\bv^j}_{t+1} &= \{ \bu^i_{t+1} \mid i \st \bv^i \in \mathcal{N}_{t+1} (\bv^j) \}
\end{align}
while in the \textsc{Forecast} setting, we simply replace the ground truth
$\bu^i_{t+1}$ with the forecast $\hbu^i_{t+1}$. We now project the ego's
embedding into the query space,
\begin{align}
    \bu^{Qj}_t &= \bW^Q \, \bu^j_t
\end{align}
and the neighbors' embeddings into the key and value space,
\begin{align}
    \bu^{Ki}_{t+1} &= \bW^K \, \bu^i_{t+1} \qquad \forall i \st \bv^i \in \mathcal{N}_{t+1} (\bv^j) \\
    \bu^{Vi}_{t+1} &= \bW^V \, \bu^i_{t+1} \qquad \forall i \st \bv^i \in \mathcal{N}_{t+1} (\bv^j)
\end{align}
The aggregated embedding $\tbu^j_{t+1}$ is then the weighted sum of the values
with a GELU activation,
\begin{align}
    \tbu^j_{t+1} &= \GELU \Big( \sum_{i} \lambda_i \, \bu^{Vi}_{t+1} \Big) \label{eq:attn_avg}
\end{align}
where the weights $\lambda_i$, called {\em attention scores},
are computed from the dot product between the query and the keys, followed by
a softmax.
Note that the ego node is not included in the aggregation; instead it is added
separately,
\begin{align}
    \hbu^j_{t+1} &= \bW^E \bu^j_t + \bW^N \tbu^j_{t+1} \label{eq:add_proj}
\end{align}
which is then projected down to $\R^D$,
\begin{align}
    \hbv^{jA}_{t+1} &= \bW^A \, \hbu^j_{t+1}
\end{align}
The vector $\hbv^{jA}_{t+1}$ is the forecast contribution from the flow
aggregation component, i.e. the second term in \cref{eq:sum}.

We call the full model with multi-head attention \textit{\modelname}. Note that
the flow aggregation component and the recurrent component are decoupled. Thus
we can easily substitute the multi-head attention with another node aggregation
method. In particular, if we replace \cref{eq:attn_avg} with a simple
arithmetic average of the neighbors,
\begin{align}
    \tbu^j_{t+1} &= \dfrac{1}{|\mathcal{N}_{t+1} (\bv^j)|} \sum_{i} \bu^{i}_{t+1} \label{eq:graphsage}
\end{align}
we would obtain the original formulation of
GraphSage~\cite{Hamilton2017GraphSAGE}. We  call the model that uses
\cref{eq:graphsage} instead of \cref{eq:attn_avg} \textit{\modelsage}. In
addition to adopting \cref{eq:graphsage}, a further simplification is to remove
the linear projection in \cref{eq:add_proj} when adding the ego's embedding
with its neighbors'. Let us call this variant \textit{\modelmean}.

\subsection{Relationship with existing models}
\label{ssec:model_discussion}

\subsubsection{GAT}

Our multi-head attention neighborhood aggregation is similar to the Graph
Attention Network (GAT)~\cite{Velickovic2018GAT}. To compute the attention
score in GAT, we first need to concatenate the ego node's embedding with the
neighbor's, and then feed the result through a single feedforward network
followed by a LeakyReLU. In contrast, we revert back to the original multi-head
attention~\cite{Vaswani2017AttentionIA} where we compute the attention scores
with a simple dot product. We also add zero attention, in which a node has the
option not to attend to any neighbor. We will empirically show in
\cref{sec:results} that our simpler method outperforms GAT in almost all
settings.

\subsubsection{N-BEATS}

The process of feeding residuals of time series into deep network layers is
inspired by N-BEATS. However, N-BEATS takes residuals from the raw scalar
observations, whereas our approach calculates the residuals from the
vector-valued projections of the time series, as shown in \cref{eq:input}.
Moreover, N-BEATS is not easily adapted to the dynamic graph setting since it
does not produce embeddings that depend on time. N-BEATS treats the forecasting
task as a multivariate regression problem, where every step can see every other
step in the history. This allows us to obtain an embedding for the whole series
but not for an individual step. In our proposed architecture, the node vector
$\bu^{j \ell}_{t}$ is used to construct the time-dependent embedding of each
step, as shown in \cref{eq:node_embed}.

\subsubsection{Transformers}

In the last few years, transformers~\cite{Vaswani2017AttentionIA} have become
the sequence model of choice in the NLP domain. Despite their success in NLP
tasks, little progress has been made with time series forecasting. Most
recently, \citet{Wu2020DeepTM} designed a transformer to forecast flu cases,
but their model provides only marginal improvement over the LSTM baseline. Our
preliminary investigation indicated that LSTMs perform better than transformers
in the time series setting. We hypothesize that the strict temporal ordering of
the LSTM can encode time series more naturally; while text, which often has a
latent tree structure, is more naturally encoded by the transformer with its
attention mechanism and position encodings.

\subsubsection{Non-neural aggregation}
\label{ssec:arnet}

The most relevant non-neural aggregation approach is
ARNet~\cite{Wu2019EstimatingAF}, a forecasting model for scalar-valued time
series in which the prediction is computed as:
\begin{align}
\hv^{j}_{t} = \sum_{k=1}^{p}
\alpha^{j}_{k} v^{j}_{t-k} + \sum_{v^i \in \N(\bv^j)} \beta^{ij} v^i_{t}
\end{align}
where the first term is an autoregressive model of order $p=7$ (days) and the
second term models the network effect. The learnable parameters $\beta^{ij}$
can be interpreted as the edge weight that controls the proportion of views
propagating from node $i$ to node $j$. Although ARNet is simple with a
straightforward interpretation, the model assumes that the network is static.
Furthermore \citet{Wu2019EstimatingAF} only evaluated on the
\textsc{Imputation} setting and not on the \textsc{Forecast} setting where the
future observations are unknown. We will show in \cref{sec:results} that the
added complexity of \modelname allows it to both incorporate dynamic graphs and
function in the \textsc{Forecast} setting.

\subsubsection{Neural aggregation}

The closest model to ours is T-GCN~\cite{Zhao2019TGCNAT}, where a modified GRU
cell does a graph convolution before computing the update and reset gates.
Unlike \modelname which has been implemented to fetch subgraphs from disk and
only compute the network information once after the final LSTM layer, T-GCN
requires the entire network to be in memory and aggregates the network at every
time step in every layer. Therefore, T-GCN does not scale to larger datasets
due to both space and time complexity. Our proposed architecture, on the other
hand, can easily handle dynamic networks of hundreds of thousands of nodes.

%% file: 07_dataset.tex
\begin{table}[t]
  \caption {Key statistics on two snapshots (the first and last day)
            of \Vevo and \Wiki.}
  \Description{Key statistics on two snapshots of the networks.}
  \label{tab:graph_stats}
  \centering
  {\fontsize{7.8}{10}\selectfont
  \begin{tabularx}{\columnwidth}{Xrrrr}
    \toprule
    & \multicolumn{2}{c}{\Vevo} & \multicolumn{2}{c}{\Wiki} \\
      \cmidrule(lr){2-3} \cmidrule(lr){4-5}
      & 1 Sep 18 & 2 Nov 18 & 1 Jul 15 & 30 Jun 20\\
    \midrule
    Number of nodes & 60,663  & 60,664  & 329,255 &  366,145                 \\
    Number of edges & 1,189,460 & 1,192,478 & 12,869,374 & 17,417,749 \\
    Nodes with in-edges & 56,852 & 56,672 & 303,440 & 356,120 \\
    \quad Mean in-degree & 21 & 21 & 42 & 49 \\
    \quad Median in-degree & 10 & 10 & 14 & 17 \\
    Nodes with out-edges & 60,553 & 60,545 & 319,426 & 362,148 \\
    \quad Mean out-degree & 20 & 20 & 40 & 48 \\
    \quad Median out-degree & 19 & 19 & 26 & 31\\
    Diameter & 32 & 27 & 15 & 22 \\
    Average path length & 8.4  & 8.3 & 4.1 & 4.0 \\
    Clustering coefficient & 0.17  & 0.15 & 0.014 & 0.015\\
    \bottomrule
  \end{tabularx}}
\end{table}

\section{Dynamic Networks of Time Series}

The empirical validation of \modelname is carried out on two small static
networks---\LA and \SZ~\cite{Zhao2019TGCNAT}; and two large-scale dynamic
networks---\Vevo~\cite{Wu2019EstimatingAF} and \Wiki. Out of these, \Wiki is a
new dataset that we have collected and it is the largest dynamic network of
time series to date. This section describes each dataset in detail.

\subsection{\LA and \SZ}

\LA and \SZ~\cite{Zhao2019TGCNAT} contain time series of traffic speeds and
road network information. \LA is a network of 207 sensors, measuring traffic
speeds at 5-minute intervals from 1 March to 7 March 2012. There is an edge
between two sensors if they are close to each other. \SZ is a network of 156
roads in the Luohu District in Shenzhen, containing 15-minute interval traffic
speeds from 1 January to 31 January 2015. If two roads are connected, an edge
is formed between them. Both are static networks, with \LA containing 2,833
edges and \SZ containing 532. We use these datasets to compare \modelname
against T-GCN~\cite{Zhao2019TGCNAT}.

\subsection{\Vevo}

\Vevo~\cite{Wu2019EstimatingAF} is a YouTube video network containing 60,740
music videos from 4,435 unique artists. Each node in the network corresponds to
a video and is associated with a time series of daily view counts collected
over the course of 63 days from 1 September 2018 to 2 November 2018. A directed
edge from video $u$ to video $v$ is present on day $t$ if $v$ appears on $u$'s
list of recommendations on day $t$.

To ensure a fair comparison, we use the chronological train-test split by
\citet{Wu2019EstimatingAF}, in which we train on the first 49 days, validate on
the next 7 days, and test on the final 7 days. We also follow the original
setup which computes evaluation metrics on the 13,710 nodes with at least one
incoming edge. This makes the differences between network and non-network
models more apparent.

\begin{figure}[t]
  \centering
  \includegraphics[width=\linewidth]{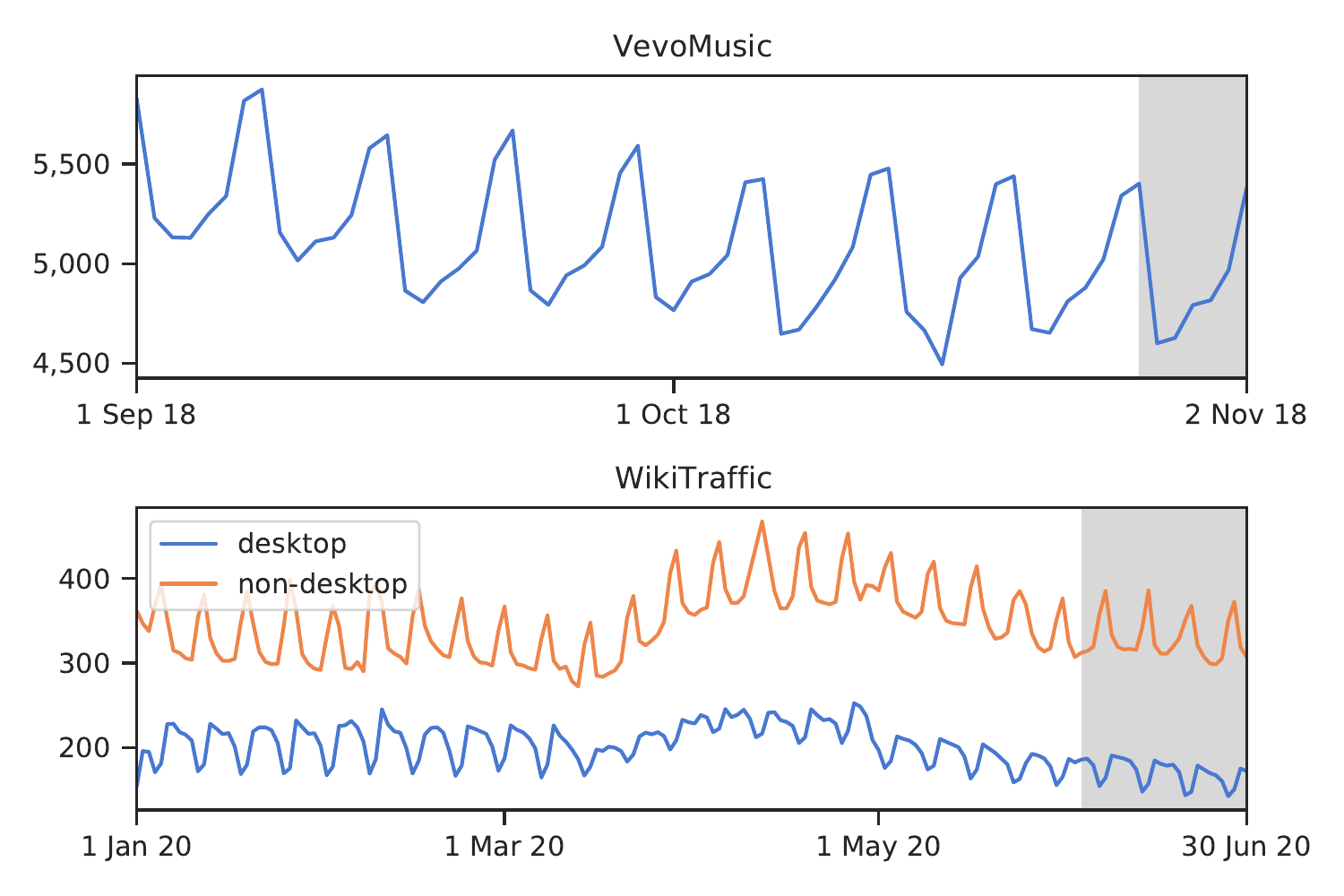}

  \caption{Ground-truth time series averaged across all samples in \Vevo\ (top)
  and \Wiki\ (bottom). Shaded areas correspond to the test period. Strong
  weekly seasonality can be observed in both datasets.}

  \Description{Time series of the average view counts.}
  \label{fig:avg_series}
\end{figure}



\subsection{\Wiki}

\begin{figure}[t]
  \centering
  \includegraphics[width=\linewidth]{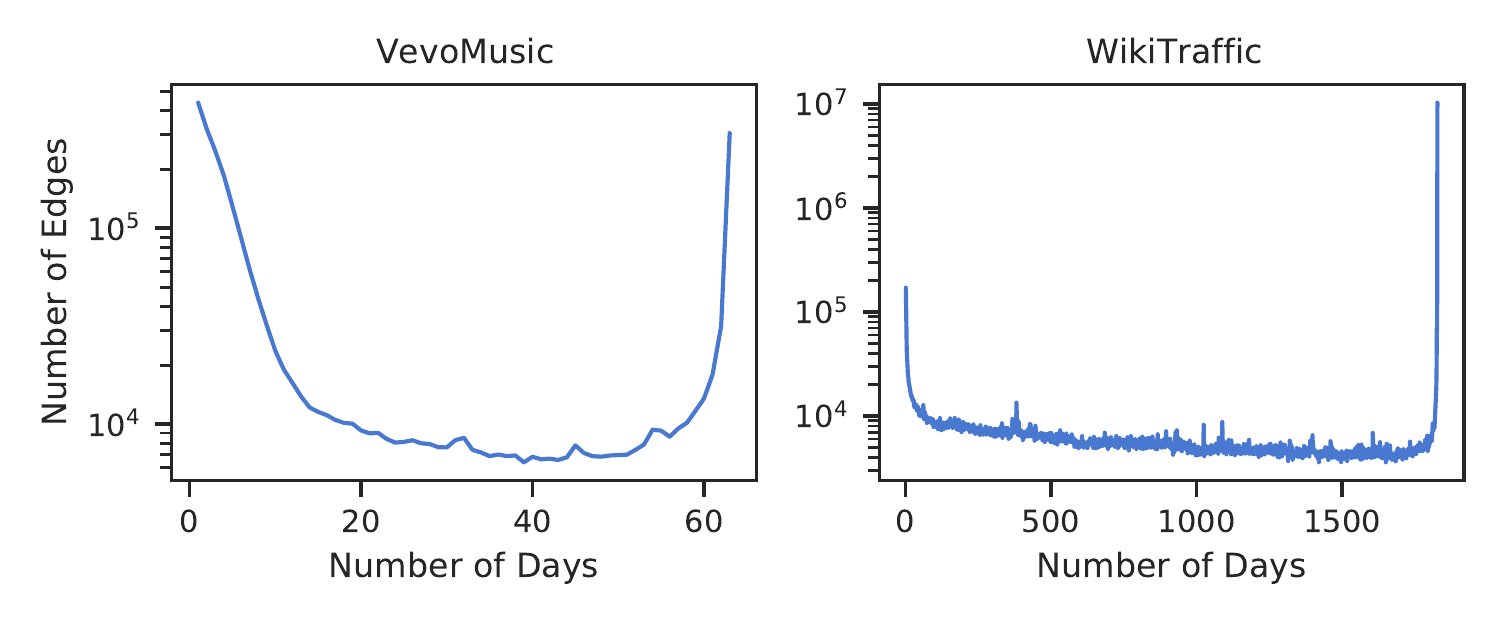}

  \caption{The distribution of link durations. In \Vevo most links are
    short-lived (50\% appear fewer than five days), likely due to YouTube
    diversifying its recommendations. In contrast, links in \Wiki are more
    stable, with half persisting throughout the entire five years.}

  \Description{The distribution of link durations.}
  \label{fig:edge_dist}
\end{figure}

We collected the new \Wiki\ network dataset which contains 366K nodes with 22
million unique page pairs that have an edge on at least one day over a
five-year period. On any given day, we have up to 17 million links, as shown
in~\cref{tab:graph_stats}. \Wiki is similar to \Vevo in that they both exhibit
a strong weekly seasonality (\cref{fig:avg_series}). They both have dynamic
links, although \Wiki links are more stable overall (\cref{fig:edge_dist}).

The data collection starts with the raw dump of the English
Wikipedia\footnote{\url{https://dumps.wikimedia.org}}
containing the full revision history of 17 million articles. From this we
collect daily view counts from 1 July 2015 to 30 June 2020. We then remove
articles with less than 100 daily average views in the final 140 days. This
leaves us with 366,802 pages. The view counts are split into two categories:
views from desktop users and from non-desktop users.
We set the final 28 days to be the test period, the 28 days before that to be
the validation period, and the remaining days for training. Forecasting 28 days
in advance allows us to test the robustness of the model when predicting a
substantial time into the future.

Furthermore, since \Wiki is an order of magnitude larger than other networked
time series datasets, we can set aside nodes to be used only during testing.
Thus the train-test split is divided both by time and by node, providing a
stronger test of a model's ability to generalize. To be useful for evaluating
network-based forecasting models, the test set should form its own network, so
we choose nodes that are connected. We start with four seed categories:
\textit{Programming Languages}, \textit{Star Wars}, \textit{Global Warming},
and \textit{Global Health}, each with many subcategories.
Starting with each seed category, we collect all pages in that category and all
subcategories within four levels. This provides 2,434 pages
for our test set. Finally we consider two versions of the dataset---a
univariate version where we predict the total view count of a page, and a
bivariate version where we predict the desktop and non-desktop traffic
separately.

Prior to our work, Google created a dataset containing two years of traffic
from 145K randomly sampled pages in Wikipedia for a Kaggle
competition\footnote{\url{https://www.kaggle.com/c/web-traffic-time-series-forecasting/data}}.
However this dataset contains no network information and includes low-traffic
pages which are noisy. \citet{rozemberczki2019multiscale} curated a small
hyperlink Wikipedia network on specialized topics (chameleons, crocodiles, and
squirrels) with only monthly view counts. \citet{consonni2019wikilinkgraphs}
introduced WikiLinkGraphs containing all dynamic links from 2001 to 2018, but
they did not collect traffic information. In contrast, our \Wiki is the largest
dynamic network of time series, enabling detailed studies of information flow
and user interests on a large scale.

%% file: 08_experiments.tex

\section{Evaluation settings}
\label{sec:evalsettings}

Evaluation is done by predicting view counts of the final 7 days on \Vevo\ and
the final 28 days on \Wiki. For \LA and \SZ, we predict the speeds in the
final hour (the final 12 steps in \LA and the final 4 steps in \SZ). As stated
in \cref{sec:problem}, we consider both the \textsc{Forecast} and
\textsc{Imputation} settings for the large datasets. On \Vevo, we evaluate on
two different networks: the full dynamic network which we call \Vevo (dynamic),
and the static network which we call \Vevo (static). To construct the static
version, \citet{Wu2019EstimatingAF} used a majority smoothing method to remove
edges that occur only briefly and made the remaining edges persistent in all
time steps. Their best model ARNet was only evaluated on this static network in
the \textsc{Imputation} setting. On \Wiki, we consider two networks: one is a
network of univariate time series of view counts, while the other is a network
of bivariate time series where desktop and non-desktop traffic are split.

Following prior forecasting work~\cite{Makridakis2000M3, Makridakis2018M4},
our main evaluation metric will be the
Symmetric Mean Absolute Percentage Error for forecast horizon $F$:
\begin{align}
  \text{SMAPE-$F$} &= \dfrac{100}{\T F D} \sum_{j=1}^{\T} \sum_{t=1}^{F} \sum_{d=1}^{D}
  \dfrac{\left| v^{j}_{td} - \hv^{j}_{td} \right|}{ \frac{1}{2}
  \left( \left| v^{j}_{td} \right| + \left| \hv^{j}_{td} \right| \right) } \label{eqn:smape}
\end{align}
where $\T$ is the number of samples in the test set, $F$ is the forecast
horizon, $D$ is the dimension of the time series, and $\hv^{j}_{td}$ is the
forecast value of the ground truth $v^{j}_{td}$.
SMAPE is interpretable with an upper bound of 200 and a lower bound of 0. It is
scale-independent, ensuring that prediction errors will be considered relative
to the magnitude of the sequence. This is important because it prevents nodes
with a large number of views from dominating the evaluation measure. A lower
SMAPE corresponds to a better fit, with it being 0 if and only if the
prediction matches the ground truth perfectly. For the two small networks of
univariate time series, \LA and \SZ, we additionally report the Root Mean
Square Error,
\begin{align}
  \text{RMSE-$F$} &= \sqrt{\dfrac{1}{\T F} \sum_{j=1}^{\T} \sum_{t=1}^{F}
  \left( v^{j}_{t} - \hv^{j}_{t} \right)^2} \label{eqn:rmse}
\end{align}
and the Mean Absolute Error,
\begin{align}
  \text{MAE-$F$} &= \dfrac{1}{\T F} \sum_{j=1}^{\T} \sum_{t=1}^{F}
  \left| v^{j}_{t} - \hv^{j}_{t} \right| \label{eqn:mae}
\end{align}
similar to what was done in the T-GCN paper~\cite{Zhao2019TGCNAT}.


\subsection{Model variants}
\label{ssec:variants}

We compare 8 time series baselines, 7 variants of networked time series,
and 7 more variants of \modelname in an ablation study.
Results of the following 8 time series baselines are in \cref{tab:nonet_smapes}
and \cref{tab:traffic}.
\begin{enumerate}
  \item[(1)] \textbf{Copying Previous Step:} We use the final observation
  before the test period as the prediction. This is the final day in \Vevo and
  \Wiki, the final 5 minutes in \LA, and the final 15 minutes in \SZ.

  \item[(2)]\textbf{Copying Previous Week:} Since we observe weekly seasonality
  in both \Vevo and \Wiki (see \cref{fig:avg_series}), a stronger baseline is
  to copy observations in the final week just before the test period and use
  them as the predictions.

  \item[(3)] \textbf{AR:} The autoregressive (AR) model used by
  \citet{Wu2019EstimatingAF} for the static \Vevo network.

  \item[(4)] \textbf{Seasonal ARIMA:} We train an ARIMA$(p,d,q)(P,D,Q)_m$ model
  separately for each time series, where $p, d, q, P, D, Q$ are the AR,
  difference, MA, seasonal AR, seasonal difference, and seasonal MA terms,
  respectively. These are learned automatically using the
  \href{https://github.com/alkaline-ml/pmdarima}{\texttt{pmdarima}} package.
  The number of periods in a season $m$ is set to 7 days for \Vevo and \Wiki.

  \item[(5)] \textbf{Individual LSTMs:} The LSTM baseline used by
  \citet{Wu2019EstimatingAF}. It is trained separately for each time series,
  with no weight sharing across network nodes.

  \item[(6)] \textbf{LSTM:} The standard LSTM model with weight sharing. Unlike
  variant (4), this method uses only one set of LSTM weights for the entire
  dataset.

  \item[(7)] \textbf{N-BEATS:} The neural regression with residual stacking
  from \citet{Oreshkin2020NBEATS}. The implementation consists of
  eight stacks, each containing one generic block. A generic block internally
  uses four fully-connected layers, followed by a fork into the forecast and
  backcast space. For bivariate \Wiki, we train two separate N-BEATS
  models.

  \item[(8)] \textbf{\modelname-NoNetwork:} \modelname with only the recurrent
  component, i.e. first term $\hat \bv^{jR}_t$ in \cref{eq:sum}. It does not
  take any contribution from the network.
\end{enumerate}
Results for the forecasting models which use the network
structure are presented in \cref{tab:traffic} and \ref{tab:net_smapes}.

\begin{table}[t]
  \caption{Hyperparameters of \textbf{\modelname-NoNetwork} (8) and \modelname
  (15) on \Wiki (univariate). We calibrate the hidden size to ensure that
  the number of parameters of all models are within 5\% of each other.
  See the
  \href{https://bit.ly/320teVC}{Appendix}~\cite{RadflowAppendix} for the
  hyperparameters of other model variants.}
  \Description{Hyperparameters.}

  \label{tab:radflow_params}
  \begin{tabularx}{\columnwidth}{Xcccc}
    \toprule
     & 0 Hops & 1 Hop & 2 Hops \\
    \midrule
    Number of parameters & 1,589,762 & 1,608,464 & 1,576,52 \\
    LSTM hidden size & 128 & 116 & 112 \\
    Feedforward hidden size & 128 & 116 & 112 \\
    Number of LSTM layers ($L$) & 8 & 8 & 8 \\
    Dropout probability & 0.1 & 0.1 & 0.1 \\
    Number of attention heads &- & 4 & 4 \\
    Backcast (seed) length &112 & 112 & 112 \\
    Forecast length &28 & 28 & 28 \\
    \bottomrule
  \end{tabularx}
\end{table}

\begin{enumerate}

  \item[(9)] \textbf{T-GCN:} The model proposed by~\citet{Zhao2019TGCNAT} that
  uses a modified GRU cell to aggregate nodes. Since T-GCN is not scalable to
  larger datasets (see \cref{ssec:model_discussion}), we only compare
  \modelname against T-GCN on \LA and \SZ~\cite{Zhao2019TGCNAT}.

  \item[(10)] \textbf{ARNet:} The state-of-the-art
  model~\cite{Wu2019EstimatingAF} for the \Vevo (static) dataset in the
  \textsc{Imputation} setting (see \cref{ssec:arnet}).

  \item[(11)] \textbf{LSTM-MeanPooling:} The same architecture as (6) but with
  mean-pooling node aggregation, using the hidden output of the final LSTM
  layer as a node's representation.

  \item[(12--15)] \textbf{\modelname:} Variants of our proposed architecture
  with different network aggregation techniques: a simple mean (12), GraphSage
  (13), Graph Attention Network (14), and our full \modelname model with
  multi-head attention (15). \cref{tab:radflow_params} outlines the
  hyperparameters used in the full model.

\end{enumerate}
Finally we conduct an ablation study to test the key components of
our architecture (\cref{tab:ablations}).
Starting with the best model (15), we substitute one
component with an alternative:

\begin{enumerate}
  \item[(16--20)] \textbf{\modelname with other node embeddings:} As shown in
  \cref{sec:radflow}, the LSTM cell contains a hidden state $\bh$ which is used
  to produce three vectors, $\bp$, $\bq$, and $\bu$. Instead of having a
  separate output $\bu$ to represent a node, we could alternatively reuse the
  cell's hidden state $\bh$ (16), the backcast representation $\bp$ (17), or
  the forecast representation $\bq$ (18). We could also concatenate different
  representations, such as $[\bh; \bp]$ (19) or $[\bh; \bp; \bq]$ (20).

  \item[(21)] \textbf{\modelname with no final projection:} We ignore the
  linear projection in \cref{eq:add_proj} and add the ego node's embedding to
  its neighbors' directly.

  \item[(22)] \textbf{\modelname with one attention head:} This final variant
  tests the effect of having only one attention head instead of the default four
  in the full model.
\end{enumerate}

\subsection{Data preprocessing and training details}

Web-scale time series observations often vary greatly in scale. An unpopular
page might get zero views, while a popular page might receive millions of
visits daily. 
To ensure similar scaling, both the inputs and outputs of our models are
log-transformed time series. Outputs are exponentiated before computing SMAPE,
RMSE, and MAE. Missing views are imputed by propagating the last valid
observation forward. We do not apply any other preprocessing techniques to the
time series, such as trend or seasonality removal.

\cref{tab:radflow_params} shows key hyperparameters of \modelname.
We trained all models on the SMAPE objective using the Adam
optimizer~\cite{Kingma2015Adam} with $\beta_1 = 0.9$, $\beta_2 = 0.999$,
$\epsilon= 10^{-8}$. We set the weight decay factor to $10^{-4}$ and decouple
it from the learning rate~\cite{Loshchilov2018Decoupled}. We warm up the
learning rate to $10^{-4}$ in the first 5,000 steps and then linearly decay it
afterward for 10 epochs, each of which consists of 10,000 steps. We clip the
gradient norm at 0.1. All our models are implemented in
Pytorch~\cite{Paszke2017Automatic}.
For a fair comparison, we
fix the number of layers of all variants to eight and ensure that size of all
variants are within 5\% of each other.

\subsection{Computational costs}

\begin{table}[t]
  \caption{Performance of time series models with no network information. We
  report mean SMAPE-7 on \Vevo\ (22 Oct 18 -- 2 Nov 18) and mean SMAPE-28 on
  \Wiki\ (3 Jun 20 -- 30 Jun 20). Rows are numbered according to
  Sec~\ref{ssec:variants}. See the Appendix~\cite{RadflowAppendix} for
  statistical significance tests.}
  \Description{Performance of time series models with no network information.}

  \label{tab:nonet_smapes}
  \small{
  \begin{tabularx}{\columnwidth}{Xccccc}
    \toprule
     &  \makecell{\textsc{Vevo}\\\textsc{Music}} &  \makecell{\Wiki \\(univariate)} & \makecell{\Wiki\\(bivariate)} \\
    \midrule
    (1) Copying Previous Step & 14.0 & 22.5 &  26.8     \\
    (2) Copying Previous Week	& 10.3 & 21.0 &  25.4     \\
    (3) AR~\cite{Wu2019EstimatingAF} & 10.2 & - &  -     \\
    (4) Seasonal ARIMA &  9.67 & 19.6 &  22.8     \\
    (5) Individual LSTMs~\cite{Wu2019EstimatingAF} & 9.99 & - &  -     \\
    (6) LSTM & 8.68 &  16.6 & 20.4      \\
    (7) N-BEATS & 8.64 & 16.6 &  20.3      \\
    (8) \modelname-NoNetwork & \textbf{8.42} &  \textbf{16.1} & \textbf{19.4}      \\
    \bottomrule
  \end{tabularx}}
\end{table}

\begin{table}[t]
  \caption{\textsc{Forecast} performance on the static traffic networks. On
  \LA, we report mean SMAPE-12, RMSE-12, and MAE-12. On \SZ, we report mean
  SMAPE-4, RMSE-4, and MAE-4.}
  \Description{Forecast performance on the static traffic networks.}
  \label{tab:traffic}
  \footnotesize{
  \begin{tabularx}{\columnwidth}{Xcccccc}
    \toprule
     &  \multicolumn{3}{c}{\LA} & \multicolumn{3}{c}{\SZ} \\
     \cmidrule(lr){2-4} \cmidrule(lr){5-7}
     &  \scriptsize{SMAPE}  & \scriptsize{RMSE} & \scriptsize{MAE} & \scriptsize{SMAPE}  & \scriptsize{RMSE} &  \scriptsize{MAE} \\
    \midrule
    (1) Copying Previous Step & 3.92 & 3.40 & 2.39   & \textbf{45.8} &  4.32 & 2.81  \\
    (8) \modelname-NoNetwork & 3.60 & 3.23  & 2.18 &  80.2 & 3.99 & 3.06 \\
    (9) T-GCN~\cite{Zhao2019TGCNAT} & 3.97 & 3.42  & 2.41 & 80.5 & 6.27 & 3.52 \\
    (15) \modelname & \textbf{3.50} & \textbf{3.11}  & \textbf{2.11} &  77.5 & \textbf{3.36}  & \textbf{2.51} \\
    \bottomrule
  \end{tabularx}}
\end{table}

\begin{table*}[t]
  \caption{Performance of models with network information. We report mean
  SMAPE-7 on \Vevo\ (22 Oct 18 -- 2 Nov 18) and mean SMAPE-28 on \Wiki\ (3 Jun
  20 -- 30 Jun 20). Rows are numbered according to Sec~\ref{ssec:variants}.
  Bold numbers indicate the best model(s) within a column. Refer to the
  \href{https://bit.ly/320teVC}{Appendix}~\cite{RadflowAppendix} for the
  p-values from the dependent t-test for paired samples between models with
  similar performance.}
  \Description{Performance of models with network information.}

  \label{tab:net_smapes}

  \small{
  \begin{tabularx}{\textwidth}{Xcccccccccccccccc}
    \toprule
    & \multicolumn{4}{c}{\Vevo\ (static)} &  \multicolumn{4}{c}{\Vevo\ (dynamic)}
    & \multicolumn{4}{c}{\Wiki\ (univariate)} &  \multicolumn{4}{c}{\Wiki\ (bivariate)} \\

    \cmidrule(lr){2-5} \cmidrule(lr){6-9} \cmidrule(lr){10-13} \cmidrule(lr){14-17}
    & \multicolumn{2}{c}{Forecast} &  \multicolumn{2}{c}{Imputation}
    & \multicolumn{2}{c}{Forecast} &  \multicolumn{2}{c}{Imputation}
    & \multicolumn{2}{c}{Forecast} &  \multicolumn{2}{c}{Imputation}
    & \multicolumn{2}{c}{Forecast} &  \multicolumn{2}{c}{Imputation} \\

    & 1H & 2H & 1H & 2H & 1H & 2H & 1H & 2H & 1H & 2H & 1H & 2H & 1H & 2H & 1H & 2H \\
    \midrule
    (10) ARNet~\cite{Wu2019EstimatingAF}
      & - & - & 9.02   & -   & -   & -   & -  & -
      & -   & -   & -  & - & -   & -   & -  & - \\

    (11) LSTM-MeanPooling
      & 8.60 & 8.67 & 8.14  & 8.13
      & 8.80   & 9.03   & 7.91  & 7.90
      & 16.8  & 16.7   & 15.5  & 15.2
      & 20.2   &  19.9   & 19.2  & 18.9 \\

    (12) \modelmean
      & 8.34 & 8.44 & 7.82  & 7.81
      & 8.42   & \textbf{8.32}   & 7.74  & 7.61
      & 16.5   & 16.5   & 15.1  & 15.1
      & 19.8   & 20.2   & 18.5  & 18.6 \\

    (13) \modelsage
      & 8.39 & \textbf{8.37} & 7.78  & 7.64
      & 8.43  & 8.46   & 7.46  &  \textbf{7.27}
      & \textbf{15.9}   & 16.7   & 14.7  & 15.0
      & \textbf{19.3}   & 19.9   & \textbf{18.2}  & \textbf{18.4} \\

    (14) \modelgat
      & 8.52 & 8.50 & 7.88  & 7.74
      & 8.43   & 8.39   & 7.44  & 7.28
      & 16.2   & \textbf{16.0}   & 15.0  & 15.2
      & 19.5   & 19.7   & 18.3  & 18.6 \\

    (15) \modelname
      & \textbf{8.33}   &  8.39   & \textbf{7.67}  & \textbf{7.63}
      & \textbf{8.37}   & 8.45   & \textbf{7.32}  & \textbf{7.27}
      &  16.2   & \textbf{16.0}   & \textbf{14.5}  & \textbf{14.8}
      & 19.9   &  \textbf{19.6}  & 18.3  & 18.5 \\

    \bottomrule
  \end{tabularx}}
\end{table*}

\subsubsection{Training time}

\Vevo experiments were trained on a Titan V GPU and \Wiki experiments were
trained on a Titan RTX GPU. The Titan RTX has twice the memory of the Titan V
and is needed to train the two-hop \modelname on \Wiki. All pure time series
models converge very quickly, taking no more than three hours to train. Models
with one-hop aggregation take up to 17 hours to train, while models with
two-hop aggregation can take up to two days. We pick the model from the epoch
with the lowest SMAPE score on the validation set as our best model.

\subsubsection{Efficient computation of graphs}

Unlike previous approaches such as T-GCN, our models do not require the whole
graph to be in memory during training. Instead we store the graph in the HDF5
format and only load one batch at a time directly from disk.

\subsubsection{Neighborhood sampling}
To keep the computation tractable, we devise an importance-based neighborhood
sampling technique. Instead of the common uniform sampling that was, for
example, used by~\citet{Hamilton2017GraphSAGE}, we propose a two-stage approach
to select neighbors. First we assign each neighbor a score
$
  \frac{\sum_d v^j_{td}}{\text{outdegree}(\bv^j_t) + 1}
$.
This score is the total view count of the neighbor at time step $t$, normalized
by the outdegree of the neighbor at that time. A self-loop is added to avoid
division by zero. Intuitively, a neighbor with a larger number of views will
have a greater influence, but the influence will be more diffuse if that
neighbor has many outlinks. Using these scores, we remove neighbors in the
bottom 10th percentile in the neighborhood of each ego node, which reduces
noise induced by aggregation.

In the second stage, we sample four neighbors during training with probability
proportional to the number of time steps that the neighbor appears in the
backcast period. During evaluation, we find using all nodes to be
computationally infeasible due to the large data size. Thus for each ego node,
we choose only the 16 most frequently appearing neighbors in the one-hop
setting and the top eight in the two-hop setting.

%% file: 09_results.tex
\section{Results}
\label{sec:results}

We first discuss prediction performances of different model variants
(\cref{ssec:results_perf,ssec:components},
\cref{tab:nonet_smapes,tab:traffic,tab:net_smapes,tab:ablations}). When
applicable, we report in parentheses the p-value (denoted as $p$) from the
dependent t-test for paired samples. All differences discussed in this section
are statistically significant. For more detailed significance tests, see the
Appendix~\cite{RadflowAppendix}. We then present a visual interpretation of
different layers in the recurrent component (\cref{ssec:layers}), followed by
insights provided by the network aggregation component
(\cref{ssec:net_contribution}). Finally, we present two preliminary studies on
the potential applications of models like \modelname: the robustness of
predictions when the network is not fully observed (\cref{ssec:missingdata}),
and the relationship between traffic surges on nodes and their attention scores
(\cref{ssec:popularnodes}).


\subsection{Forecasting and imputation performance}
\label{ssec:results_perf}
\subsubsection{{\sc Forecasting} without networks}

\cref{tab:nonet_smapes} summarizes the comparison between \modelname-NoNetwork
and the corresponding time series forecasting baselines (1--7).
The LSTM variant (6) outperforms both AR (3) and
Seasonal ARIMA (4), showing the robust performance of flexible
neural models. Furthermore, it also
outperforms models trained on individual time series (3, 4, 5),
highlighting the advantage of using large amounts of training data.
N-BEATS (7) outperforms LSTM (6) by a small margin
of 0.04 SMAPE ($p = \scnum{6e-3}$)
on \Vevo, while \modelname-NoNetwork outperforms all other baselines, showing
promises in combining the recurrent structure with the residual stacking idea
in our architecture.
Additionally,
\modelname-NoNetwork\ outperforms ARNet (10), the state-of-the-art for \Vevo
that uses network information (\cref{tab:net_smapes}), indicating that having
the right model outweighs having more information for this task.


\begin{table}[t]

  \caption{Ablation study on the key components of \modelname on one-hop
           \Vevo networks. See \cref{ssec:components}.}
  \Description{Ablation study on the key components of the model.}

  \label{tab:ablations}
  \small{
  \begin{tabularx}{\columnwidth}{Xccccc}
    \toprule
     &  \makecell{\Vevo \\ (static)} &  \makecell{\Vevo \\(dynamic)} \\
    \midrule
    (15) \modelname & \textbf{7.67} & \textbf{7.32} \\
    (16) \modelname ($\bh$ as embeddings) & 7.78 & 7.51 \\
    (17) \modelname ($\bp$ as embeddings) & 7.77 & 7.49 \\
    (18) \modelname ($\bq$ as embeddings) & 7.84 &  7.35 \\
    (19) \modelname ($[\bh; \bp]$ as embeddings) & 7.75 &  7.39 \\
    (20) \modelname ($[\bh; \bp; \bq]$ as embeddings) & 7.76 &  7.38 \\
    (21) \modelname (no final projection) & 7.80 & 7.33 \\
    (22) \modelname (one attention head) & 7.77 & 7.43 \\
    \bottomrule
  \end{tabularx}}
\end{table}

\subsubsection{{\sc Forecasting} with networks}

\cref{tab:traffic,tab:net_smapes}
summarize performances of models (9--15) on the four networked time series
datasets. Our full model (15) outperforms T-GCN in \LA by a non-trivial margin
on all metrics. \SZ is more noisy and no model is able to beat the SMAPE from
copying the previous step. This is because \SZ contains many consecutive zero
measurements, which the copying baseline is able to take advantage of. On
non-zero test measurements, \modelname is able to outperform all other
variants. See the Appendix~\cite{RadflowAppendix} for these results.

Across all eight {\sc Forecasting} settings on \Vevo and \Wiki, the
top-performing models are all \modelname variants.
Compared to \modelname-NoNetwork (8), incorporating one-hop neighbors improves
the SMAPE score on \Vevo from 8.42 to 8.33 ($p = \scnum{2e-33}$). On \Wiki
(univariate), using one-hop GraphSage (13) improves SMAPE from 16.1 to 15.9
($p = \scnum{2e-21}$). This confirms the recently reported network effects in
the YouTube and Wikipedia traffic~\cite{Wu2019EstimatingAF, zhu2020content}.
Note that \Wiki is much larger and more diverse, making consistent improvements
harder, hence the smaller effect size compared to \Vevo.

Finally, compared to the smoothed static network, using the dynamic network in
\Vevo only improves the performance for some model variants.
This is because the static graph was constructed from the smoothing
of edges (i.e. uncommon edges were removed), which reduces noise in the absence
of ground-truth views from the neighbors. More generally, although \modelname
is designed to handle dynamic graphs, it is impossible to know a priori whether
dynamic edges will improve the prediction performance in a given data domain.

\subsubsection{\textsc{Imputation}}

Having ground-truth observations of neighboring nodes during the forecast
period leads to substantially better SMAPE scores for all models. Models that
perform well in the \textsc{Forecast} setting also perform well in the
\textsc{Imputation} setting, with \modelname (15) achieving the highest SMAPE
in six of the eight \textsc{Imputation} settings. The performance gain going
from one hop to two hops is more evident in the \textsc{Imputation} setting.
For example, there is a boost from 7.67 to 7.63 ($p = \scnum{1e-8}$) on the
static \Vevo, and from 7.32 to 7.27 ($p = \scnum{3e-14}$) on the dynamic \Vevo.
Compared to the previous state-of-the-art ARNet~\cite{Wu2019EstimatingAF}, our
best model achieves a SMAPE score that is 19\% better (from 9.02 to 7.27).

On \Wiki, using two-hop neighbors uniformly lowers the performance, while the
training time more than doubles. Compared to \Vevo, each Wikipedia page has
many more links, most of which would be ignored by the reader. The network
effect in \Wiki is thus weaker than \Vevo (see
\cref{fig:network_contribution}), leading to more noise being introduced in the
two-hop setting.


\begin{figure}[t]
  \centering
  \includegraphics[width=\linewidth]{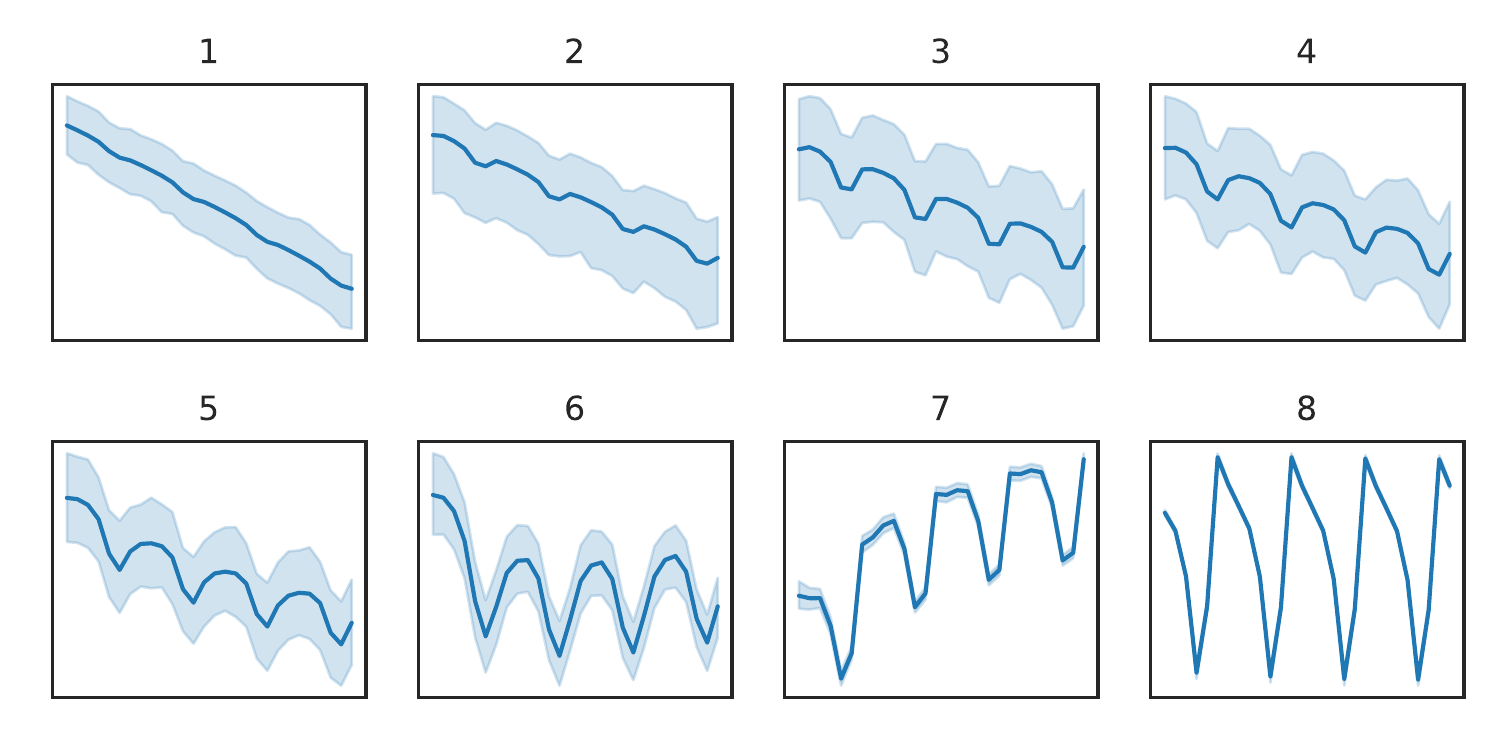}

  \caption{Average prediction over all test pages from each layer of
  \modelname-NoNetwork on the 28-day test period of \Wiki. The light shade
  corresponds to the 95\% confidence interval. Different layers capture
  different seasonal and trend patterns. See \cref{ssec:layers}.}

  \Description{The layer-wise decomposition of the time series predictions.}
  \label{fig:decomposition}
  \vspace{-1.7em}
\end{figure}

\subsection{\modelname components}
\label{ssec:components}

Among the variants with attention, the weighted multi-head attention of
\modelname yields the best performance in 10 out of 16 tasks across {\sc
Forecast} and {\sc Imputation}, while GraphSage, which puts an equal weight on
all neighbors, is the second best (best in 6 out of 16 tasks). This indicates
that simpler attention mechanisms (inner-product attention in \modelname and
node averaging in \modelname-GraphSage) are preferred over more complex ones
(\modelname-GAT), but it should not be too simple (\modelname-MeanPooling).


\cref{tab:ablations} presents an ablation study that tests other key
components in \modelname. Overall, we find that having more than one attention
head helps (15 vs 22), so does a linear projection before node aggregation (15
vs 21). It is also preferable to have a separate projection on the output of
the LSTM cell to obtain the node embeddings $\bu$, than to re-use the hidden
state $\bh$, the backcast representation $\bp$, or the forecast representation
$\bq$ (15 vs 16--20). All differences are statistically significant.

\subsection{Layered decomposition of time series}
\label{ssec:layers}


The recurrent component of \modelname is decomposable into $L=8$ layers, via
\cref{eq:sum_of_q,eq:vhat_from_qhat}. \cref{fig:decomposition} shows the
layer-wise contribution to the forecast from \modelname-NoNetwork. The results
are averaged over all 2,434 test pages in \Wiki over the 28-day test period,
and re-scaled to the same range for readability. We observe that component 8
encodes strong weekly seasonality that is consistent across all test pages
(with a small confidence interval). Components 1--5 encode varying levels of a
decreasing trend, whereas component 7 encodes an increasing trend. Overall,
weekly seasonality is visible in all components except the first, confirming
the common intuition that representations learned via neural networks are often
over-complete.

\begin{figure}[t]
  \centering
  \includegraphics[width=\linewidth]{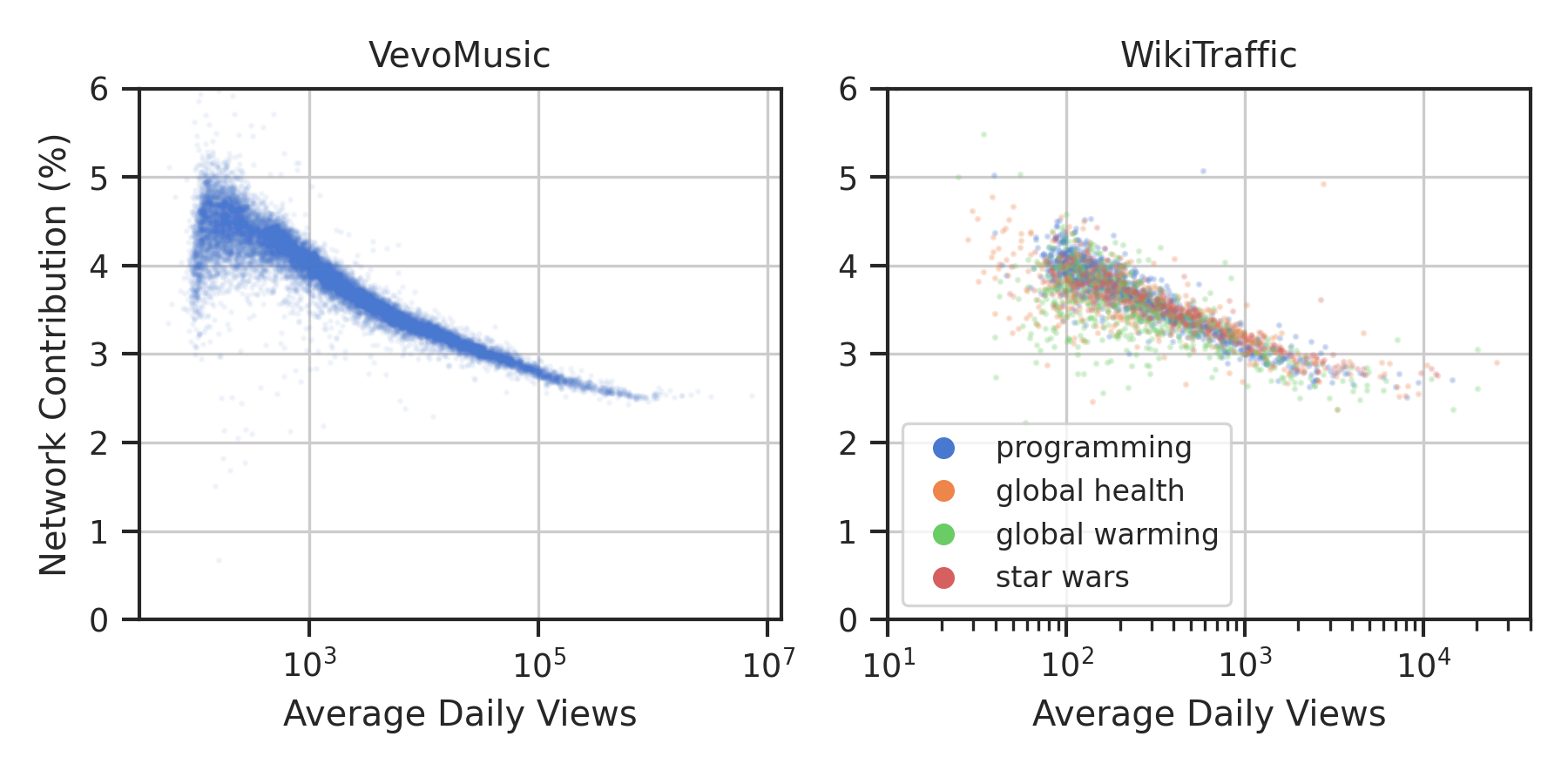}

  \caption{The contribution from the network on the final forecasts.
  The network component of less popular nodes is larger. For nodes with similar
  daily view counts, \Vevo (left) exhibits a stronger network effect than \Wiki
  (right). See \cref{ssec:net_contribution}.}

  \Description{The contribution from the network on the final forecasts.}
  \label{fig:network_contribution}
\end{figure}

\subsection{Quantifying effects of the network}
\label{ssec:net_contribution}

\subsubsection{Network contribution}
\label{ssec:v_net}

\modelname can identify settings where the network information becomes
important for the final predictions. \cref{fig:network_contribution} shows the
contribution to the final forecast made by the network component (the second
term of \cref{eq:sum}). In both \Vevo and \Wiki, there is an approximately
inverse linear relationship between the network contribution and the log of
average daily views, indicating that forecasts of less popular nodes rely more
heavily on the network.
For nodes with a similar number of daily views (take, for example, $10^3$ views
per day), \Vevo exhibits a stronger network effect ($\sim$4\%) than \Wiki
($\sim$3\%), indicating that user behaviors on YouTube recommender links and
Wikipedia hyperlinks are different, warranting further investigation (e.g.,
with user-level data).

\subsubsection{Visualising attention flow}
\label{ssec:demo}
The attention scores in \modelname capture some information about the flow of
traffic. In particular, we observe that the model pays more attention to a
neighboring node on days that have a spike in traffic (see the
Appendix~\cite{RadflowAppendix} for a specific example). This motivates us to
visualize the attention flow among \Wiki nodes in an interactive web
app~\cite{Shin2021AttentionFlow}. In this visualization,
nodes are represented in a graph (bottom panel) and as time series (top
panel). Edge weights are attention scores from \modelname multiplied by the
traffic on the source node.

\begin{figure}[t]
  \centering
  \includegraphics[width=\linewidth]{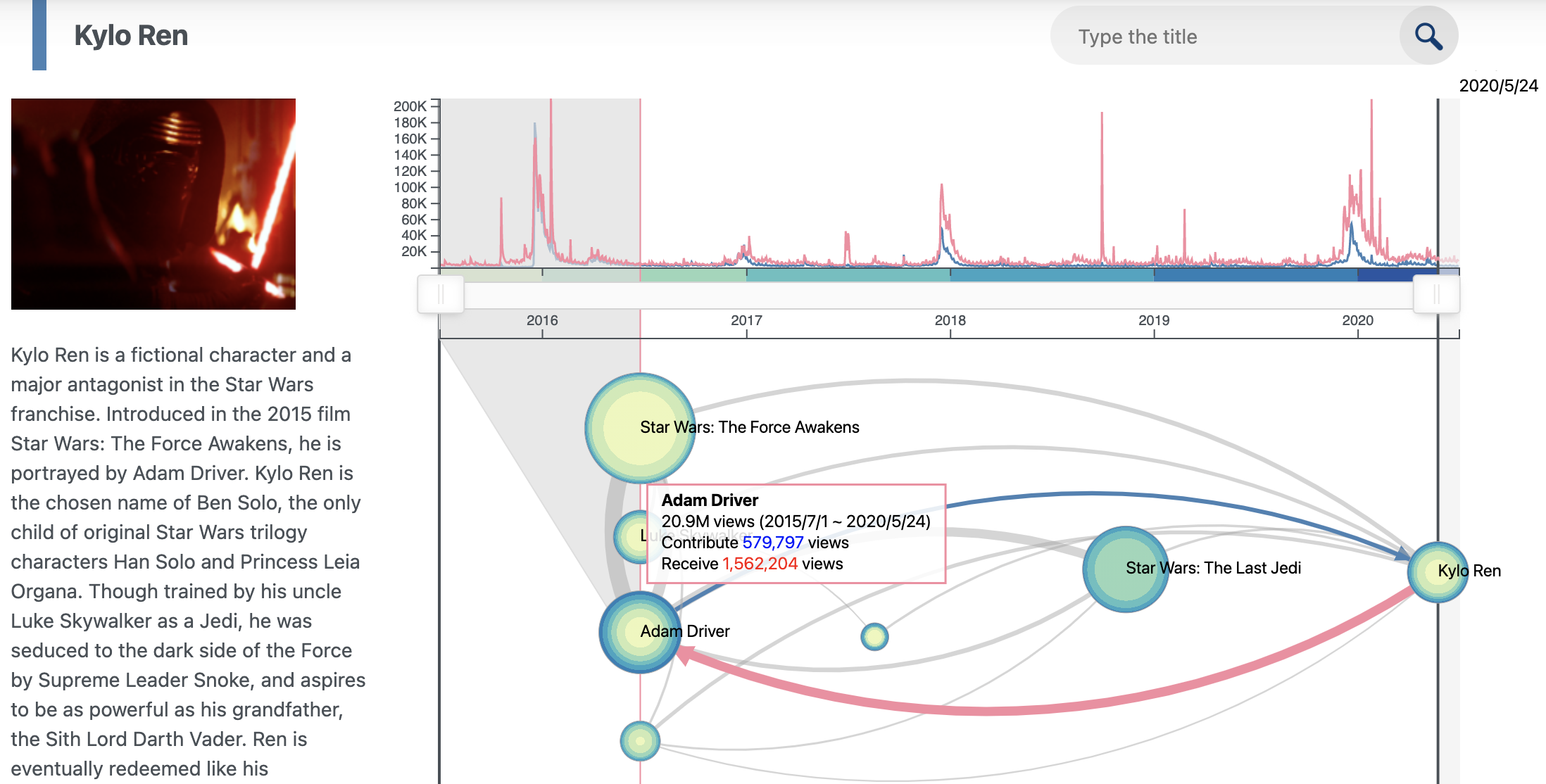}

  \caption{An interactive web app~\cite{Shin2021AttentionFlow} for attention
  flow, available at \url{https://attentionflow.ml}. Shown here is the subgraph
  for the Wikipedia page \textit{Kylo Ren}, a character in \textit{Star Wars}
  played by \textit{Adam Driver}. The thickness of an edge is determined by the
  product between the attention score and the traffic volume of the source
  node. See \cref{ssec:demo}.
  }
  \Description{An interactive web app for attention flow.}
  \label{fig:demo}
\end{figure}

\cref{fig:demo} is a screenshot for the subgraph centered around the Wikipedia
page of {\it Kylo Ren}, a character in the Star Wars series played by the actor
{\it Adam Driver}.
From the figure, we observe that the time series of the two nodes ({\it Kylo
Ren} in blue, {\it Adam Driver} in pink) have synchronized spikes at the same
time as the release of major Star Wars movies. Furthermore, there appears to be
more traffic flowing from \textit{Kylo Ren} to \textit{Adam Driver} than the
other way, indicated by the thickness of the edges.

\subsection{What if \ldots}
\label{ssec:whatif}

\subsubsection{Data is missing?}
\label{ssec:missingdata}

\begin{figure}[!t]
  \centering
  \includegraphics[width=\linewidth]{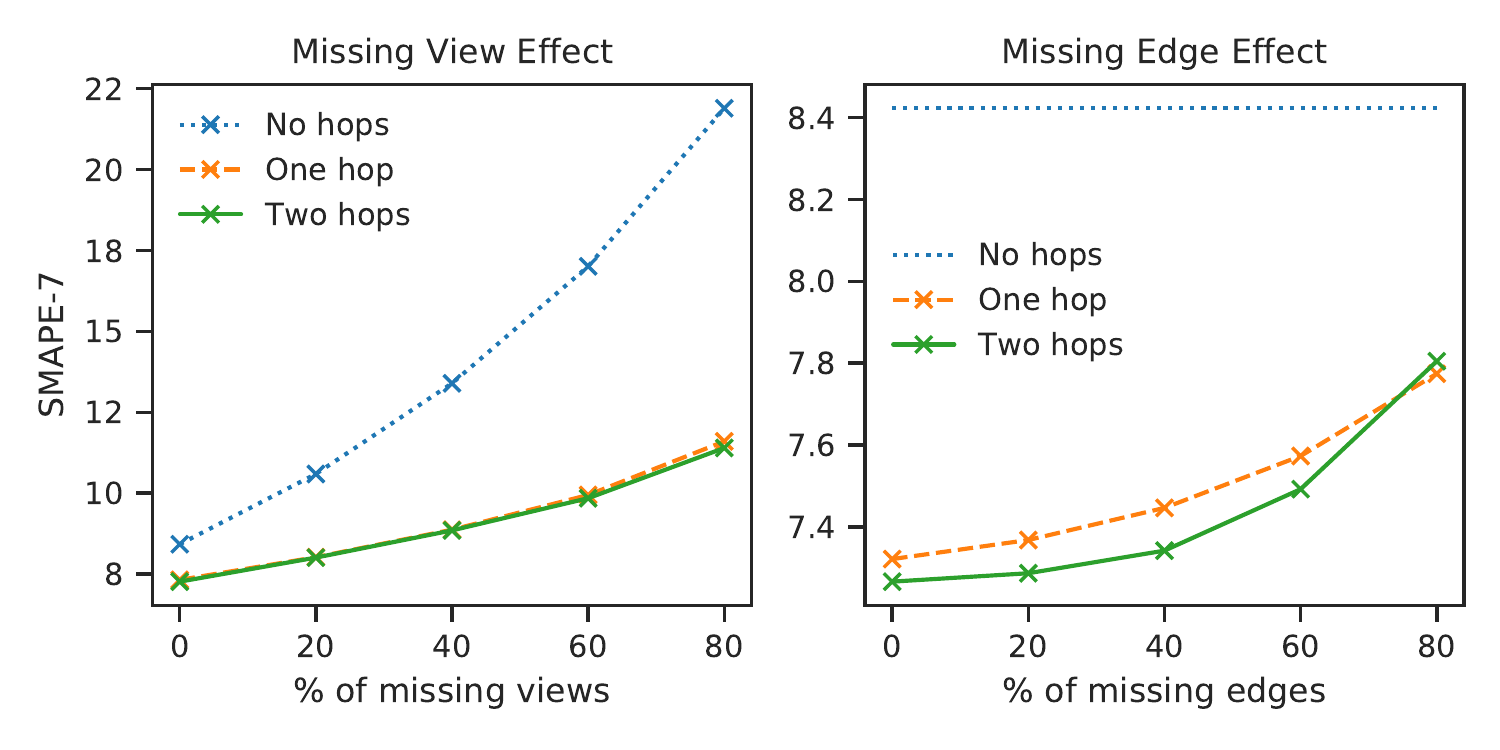}

  \caption{The effect of missing view counts (left) and missing edges (right)
           on \Vevo using \modelname. As we delete more data, \modelname's
           performance degrades at a much slower rate than \modelname-NoNetwork
           (in blue). See \cref{ssec:missingdata}.
           }

  \Description{The effect of missing view counts and missing edges on the
  performance.}
  \label{fig:missing_views}
\end{figure}

We consider the robustness of \modelname to missing data. This setting is
relevant when collecting and observing data from all nodes is very costly
(e.g., sites spread out over large geographical areas) or when nodes are simply
unavailable (e.g., sites whose data are proprietary). To this end,
\cref{fig:missing_views} shows our evaluation of \modelname\ on the {\sc
imputation} task for \Vevo with a percentage of time series values (left pane)
or edges (right pane) deleted at random. As more nodes become missing, the
performance of \modelname decays at a much slower rate than
\modelname-NoNetwork.
This indicates that \modelname is effective in imputing and mitigating missing
values from other neighbors.
Similarly, \modelname\ is relatively robust to missing edges. With 40\% of
edges missing, the performance of the two-hop model only drops by 1\% in SMAPE.
Even with 80\% of edges missing, \modelname is still substantially better than
\modelname-NoNetwork.

\begin{figure}[!t]
  \centering
  \includegraphics[width=\linewidth]{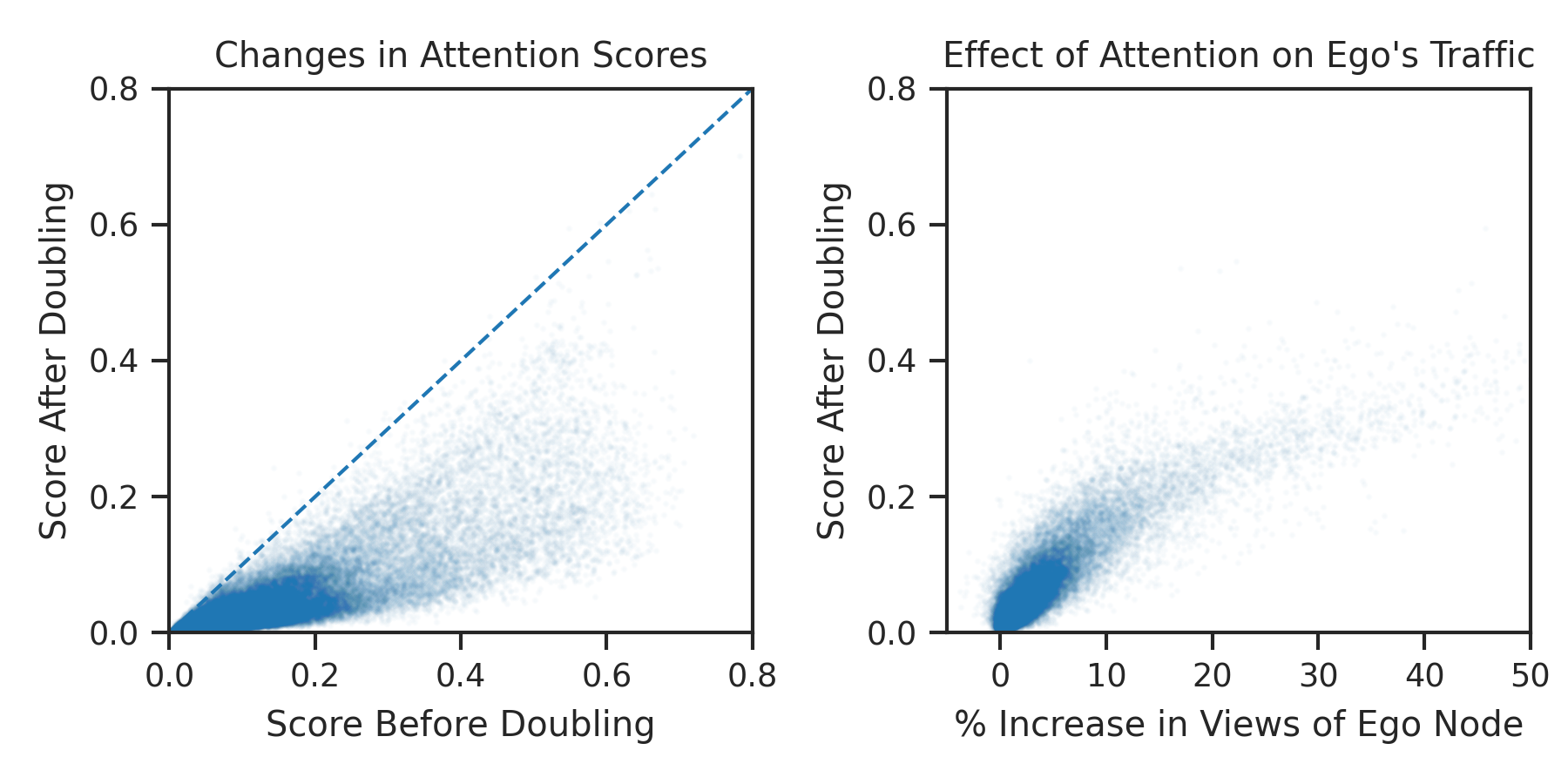}

  \caption{The effect of doubling a neighbor's view count during the forecast
  period. Left: scatter plot of attention scores before (x-axis) and after
  (y-axis) doubling. Right: Scatter plot of the relative increase of the ego
  node's views (x-axis) and attention scores after doubling (y-axis). Attention
  scores decrease as a neighbor becomes more popular (left). A higher attention
  corresponds to a larger flow of traffic from the neighbor to the ego node
  (right). See \cref{ssec:popularnodes}.}

  \Description{The effect of doubling neighbor's views during the forecast
  period.}
  \label{fig:counterfactual}
\end{figure}

\subsubsection{Node traffic doubles?}
\label{ssec:popularnodes}
In addition to the example of an actual traffic spike in \cref{fig:teaser}, we
further perform an evaluation to visualize the effects of hypothetical sudden
changes in node traffic. This scenario is useful for planning resources such as
edge network caching for mobile videos, or for estimating economic demands
associated with nodes such as advertising.

For each test node in the {\sc imputation} task on \Wiki, we pick one neighbor
and double its views on one forecasting day. \cref{fig:counterfactual} (left)
shows that as a page becomes more popular, the attention on it uniformly
decreases, indicating that attention tends to dampen node traffic spikes
instead of amplifying them.
More evidently, we can see from \cref{fig:counterfactual} (right) that a higher
attention score {\em is positively correlated with 
a large effect on the ego node's traffic}. This indicates that despite the
non-linear relationship between attention scores and network component in
\cref{sec:radflow}, one could qualitatively infer the traffic flow from one
page to another using attention scores.

%% file: 10_conclusion.tex

\section{Conclusion}

We propose \modelname, an end-to-end model for forecasting a network of time
series. It is expressive with a stack of recurrent units representing different
components of the time series, scalable to hundreds of thousands of nodes in a
network using multi-head attention and importance-sampling on neighbors, able to
represent underlying networks that change over time, and suitable for
multivariate networked series with missing nodes and edges. We achieve
state-of-the-art results on recent web-scale networked time series. We also
show that the stack of recurrent units successfully decomposes time series into
different seasonal and trend effects, and that the network attention aggregates
and explains influence between nodes. Future work includes extending \modelname
to other networked data types such as
events over
continuous time. One can explore a wide range of applications such as imputing
geographic data or allocating network resources. It would also be interesting
to investigate causal reasoning or counterfactual modeling with \modelname-like
structures.


%% file: appendix.tex
\appendix


\section{On \LA and \SZ}
\label{apdx:traffic}

\begin{figure}[b]
  \centering
  \includegraphics[width=\linewidth]{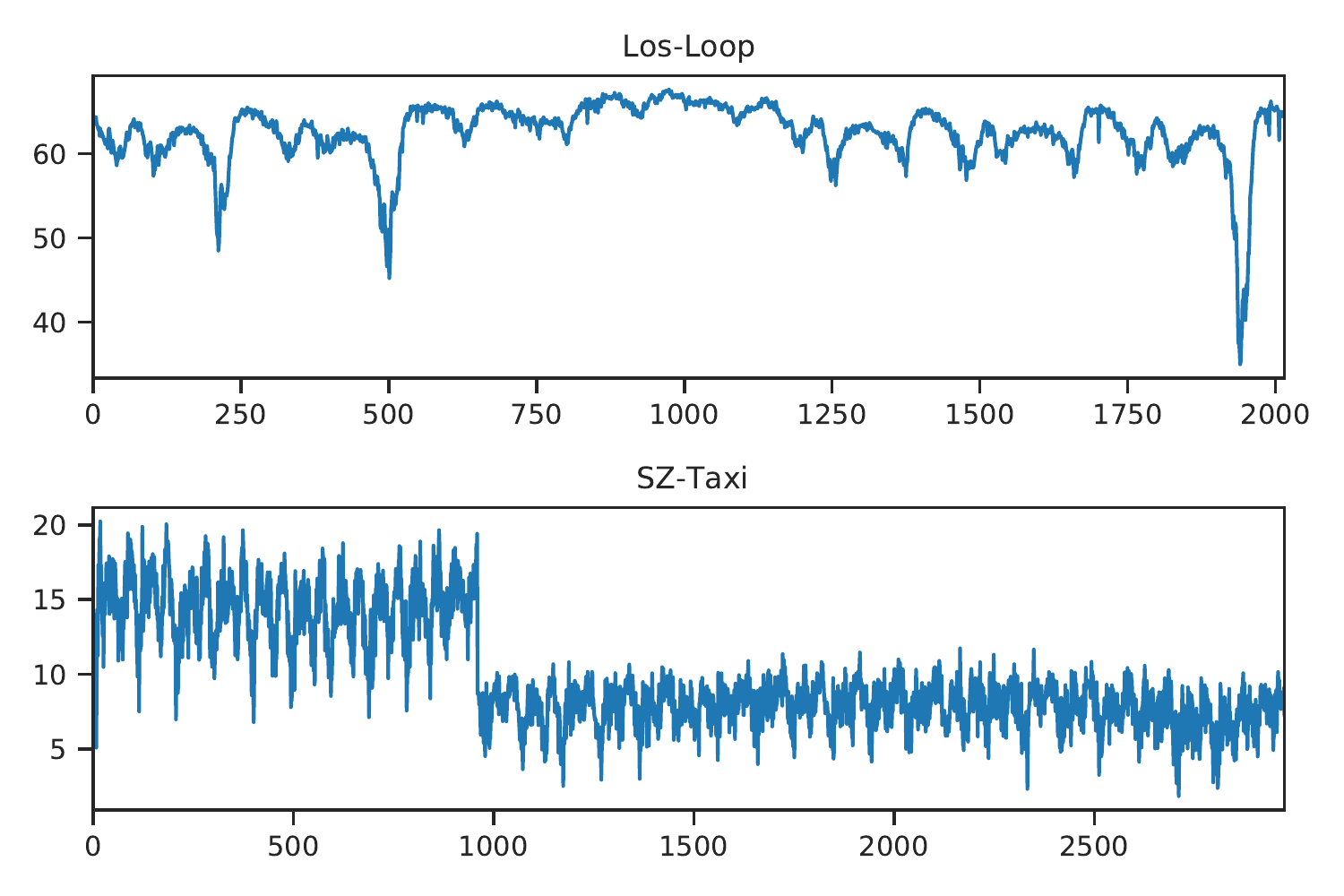}

  \caption{The median of the ground-truth time series across all samples in
  \LA\ (top) and \SZ\ (bottom).}

  \Description{The median of the ground-truth time series in the urban traffic
  datasets.}
  \label{fig:taxi_series}
\end{figure}

\begin{figure}[b]
  \centering
  \includegraphics[width=\linewidth]{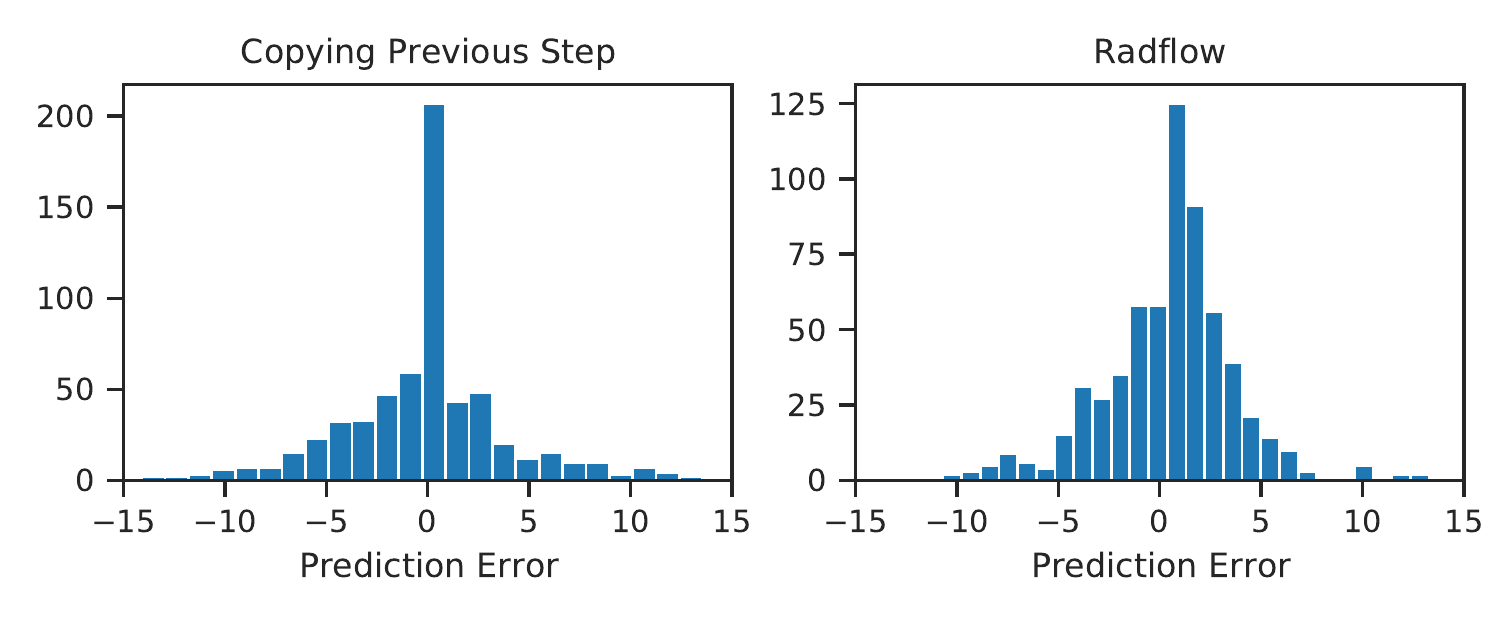}

  \caption{Prediction errors on \SZ.}

  \Description{The distribution of prediction errors.}
  \label{fig:taxi_errors}
\end{figure}

We first discuss some peculiarities found in \SZ. \cref{fig:taxi_series} shows
that \SZ is noisier than \LA, which explains \SZ's worse SMAPE scores across
all model variants in \cref{tab:traffic}. We also see a drop in the median
speed during the training period of \SZ (\cref{fig:taxi_series} bottom). This
leads to a small positive bias in the final predictions of \modelname during
the test period (\cref{fig:taxi_errors}). Such bias is not observed in the
copying baseline.

Upon closer inspection of the ground-truth measurements in \SZ's test period,
we find that 28\% of the data points are exactly zero. Many of these zero
measurements happen consecutively. Recall that the SMAPE metric is sensitive to
zero values. In particular, if the ground-truth value is zero and the predicted
value is non-zero, we obtain the maximum SMAPE of 200. Meanwhile, both RMSE and
MAE do not suffer from this problem at zero. This explains why in
\cref{tab:traffic}, the \modelname variant gets a relatively bad SMAPE score
but a good RMSE and MAE. To confirm this effect, \cref{tab:shenzhen_zero}
presents metrics both on all test measurements and on only non-zero test
measurements. We see that when we exclude the zero measurements, SMAPE
correlates better with RMSE and MAE, and \modelname now outperforms the copying
baseline.

Finally from \cref{tab:pvalues_taxi}, we see that on both datasets, our
\modelname-NoNetwork have statistically similar performance to T-GCN. Our full
\modelname outperforms T-GCN in \LA while it has comparable performance to
T-GCN in \SZ.

\begin{table}[t]
  \caption{\textsc{Forecast} performance on \SZ. We report mean SMAPE-4,
  RMSE-4, and MAE-4. We consider both the setting where we use all test
  measurements and the one where we only use non-zero measurements.}
  \Description{Forecast performance on SZ-Taxi.}
  \label{tab:shenzhen_zero}
  \footnotesize{
  \begin{tabularx}{\columnwidth}{Xcccccc}
    \toprule
     &  \multicolumn{3}{c}{All Measurments} & \multicolumn{3}{c}{Non-zero Measurements} \\
     \cmidrule(lr){2-4} \cmidrule(lr){5-7}
     &  \scriptsize{SMAPE}  & \scriptsize{RMSE} & \scriptsize{MAE} & \scriptsize{SMAPE}  & \scriptsize{RMSE} &  \scriptsize{MAE} \\
    \midrule
    (1) Copying Previous Step & \textbf{45.8} &  4.32 & 2.81  & 49.6 &  4.92 & 3.68  \\
    (8) \modelname-NoNetwork & 80.2 & 3.99 & 3.06 &  32.7 & 3.99 & 3.03 \\
    (9) T-GCN~\cite{Zhao2019TGCNAT} & 80.5 & 6.27 & 3.52 & 33.1 & 4.98 & 3.15 \\
    (15) \modelname & 77.5 & \textbf{3.36}  & \textbf{2.51}  &  \textbf{29.0} & \textbf{3.40}  & \textbf{2.52} \\
    \bottomrule
  \end{tabularx}}
\end{table}

\begin{table}[t]
  \caption{Selected p-values from the dependent t-test for paired samples on
  \LA and \SZ. The models are numbered according to Sec~\ref{ssec:variants}. We
  only show model pairs where the p-value is at least 0.001. For \SZ, we
  distinguish between the metrics computed on all test measurements
  (\SZ-\textsc{a}) and on only non-zero measurements (\SZ-\textsc{n}).}
  \Description{Selected p-values on the urban traffic datasets.}
  \label{tab:pvalues_taxi}
    \footnotesize{
      \setlength\tabcolsep{3.8pt}
  \begin{tabularx}{\columnwidth}{Xlllrlrr}
    \toprule
    \multirow{2}{*}{Dataset}
    & \multicolumn{2}{c}{Group 1} &  \multicolumn{2}{c}{Group 2} & \multirow{2}{*}{\scriptsize{p-value}} \\
    \cmidrule(lr){2-3} \cmidrule(lr){4-5}
    & Model & \scriptsize{SMAPE} & Model  & \scriptsize{SMAPE} & \\
    \midrule
    \LA   & (1) Copying Previous Step & 3.92 & (9) T-GCN & 3.97 & 0.674 \\
    \LA   & (8) \modelname-NoNetwork & 3.60 & (9) T-GCN & 3.97 & 0.002 \\
    \SZ-\textsc{a} & (8) \modelname-NoNetwork & 80.2 & (9) T-GCN & 80.5 & 0.949 \\
    \SZ-\textsc{a} & (9) T-GCN & 80.5 & (15) \modelname & 77.5 & 0.509 \\
    \SZ-\textsc{n} & (8) \modelname-NoNetwork & 32.7 & (9) T-GCN & 33.1 & 0.874 \\
    \SZ-\textsc{n} & (9) T-GCN & 33.1 & (15) \modelname & 29.0 & 0.090 \\
    \bottomrule
  \end{tabularx}}
\end{table}

\section{Further Discussion on \modelname}
\label{apdx:vevo_wiki}

We start by discussing possible interpretations of the attention scores in
\modelname. We then present results on how the performance is affected by the
topic category and the popularity of a page in \Wiki. Finally we provide
further hyperparameters and statistical significance tests for experiments on
both \Vevo and \Wiki.

\subsection{Attention scores}

Analyzing attention scores can provide insights into both the data and the
model. To this end, we show how two types of information are captured by the
scores---the flow of traffic from neighboring pages and the time series
correlation.

\subsubsection{Spikes from neighbors}
\begin{figure}[t]
  \centering
  \includegraphics[width=\linewidth]{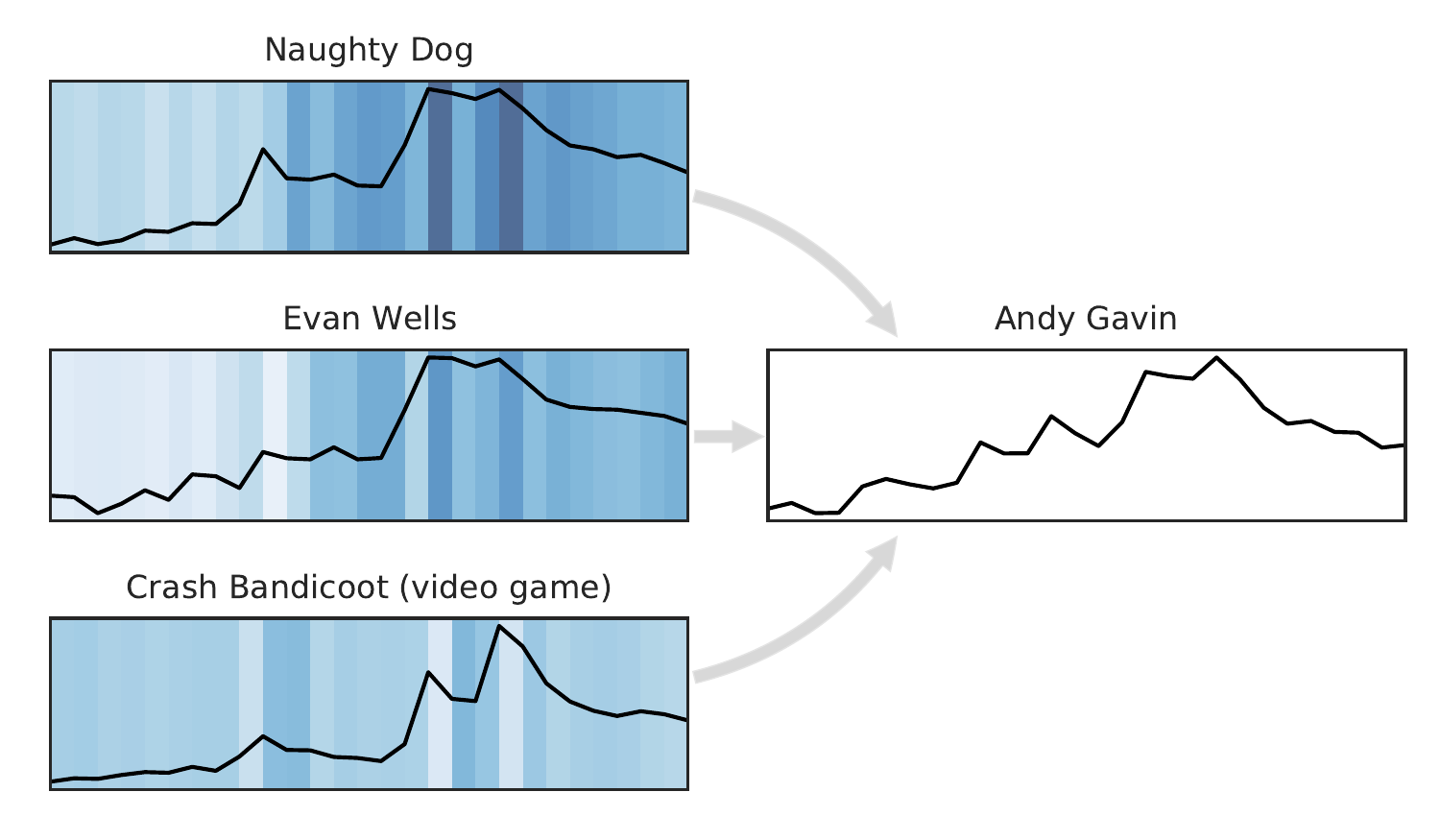}

  \caption{An example where \modelname\ attends to three neighbors as it
           forecasts the traffic of
           \href{https://en.wikipedia.org/wiki/Andy_Gavin}{\textit{Andy Gavin}}
           (video game programmer and entrepreneur). A darker blue corresponds
           to a higher attention on the corresponding day. The network effect
           is important when there is a surge in traffic.}

  \Description{An example of how attention is distributed amongst neighbors.}
  \label{fig:attention}
\end{figure}

\cref{fig:attention} shows view counts of Andy Gavin's (a video game
programmer) Wikipedia page and three linked articles with shading indicating
attention scores, i.e. $\lambda$ in \cref{eq:attn_avg}. The scores are
extracted from \modelname in the \textsc{Imputation} setting. During the
forecast period, details of a new video game \textit{Crash Bandicoot 4} were
released, leading to a spike in traffic on Andy Gavin's page (the designer of
the original \textit{Crash Bandicoot}). Linked articles, e.g., \textit{Naughty
Dog} (the company Gavin co-founded), exhibit similar spikes to which
substantial attention is applied. This example supports the intuition that
network attention is important when an exogenous event causes traffic on
neighboring nodes to change rapidly.

\subsubsection{Time series correlation}
\label{ssec:corr_attention}

\begin{figure}[t]
  \centering
  \includegraphics[width=\linewidth]{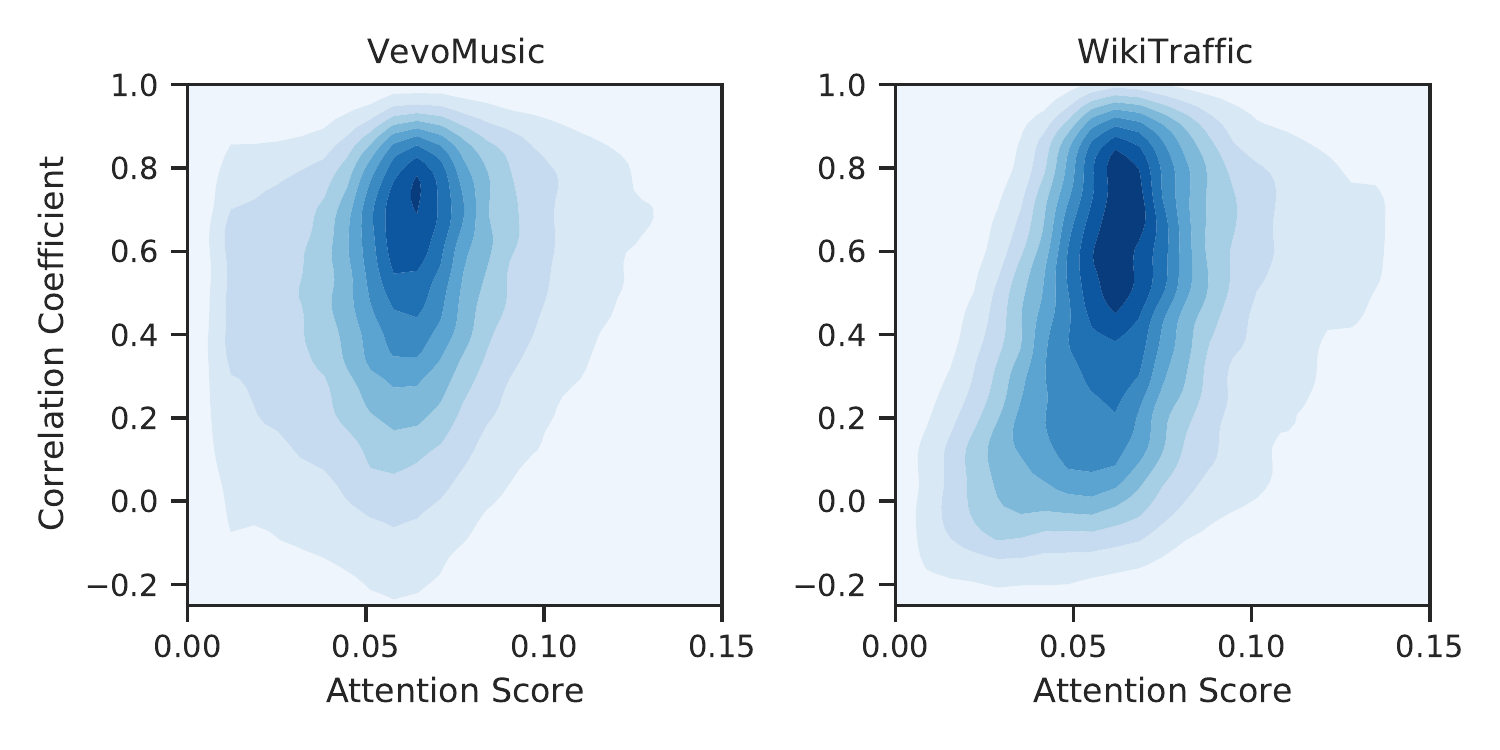}

  \caption{Density map of network attention scores (x-axis) and the correlation
  coefficient of a node with its neighbor (y-axis).}

  \Description{The density map of network attention scores and the correlation
  coefficient of a node with its neighbor.}
  \label{fig:corr_attn}
\end{figure}

To further investigate what kind of information the attention scores capture,
\cref{fig:corr_attn} presents a density plot of correlation coefficients
between the time series of an ego node and that of a neighbor, against the
average attention score on the neighbor during the forecast period. On both
\Vevo and \Wiki, we observe that if two time series have a very low to negative
correlation, the model will almost never output a high attention score. On
\Wiki, we also see that positively correlated time series rarely result in a
low attention score.

\subsection{Effect of Page Category and Popularity}

\begin{figure}[t]
  \centering
  \includegraphics[width=\linewidth]{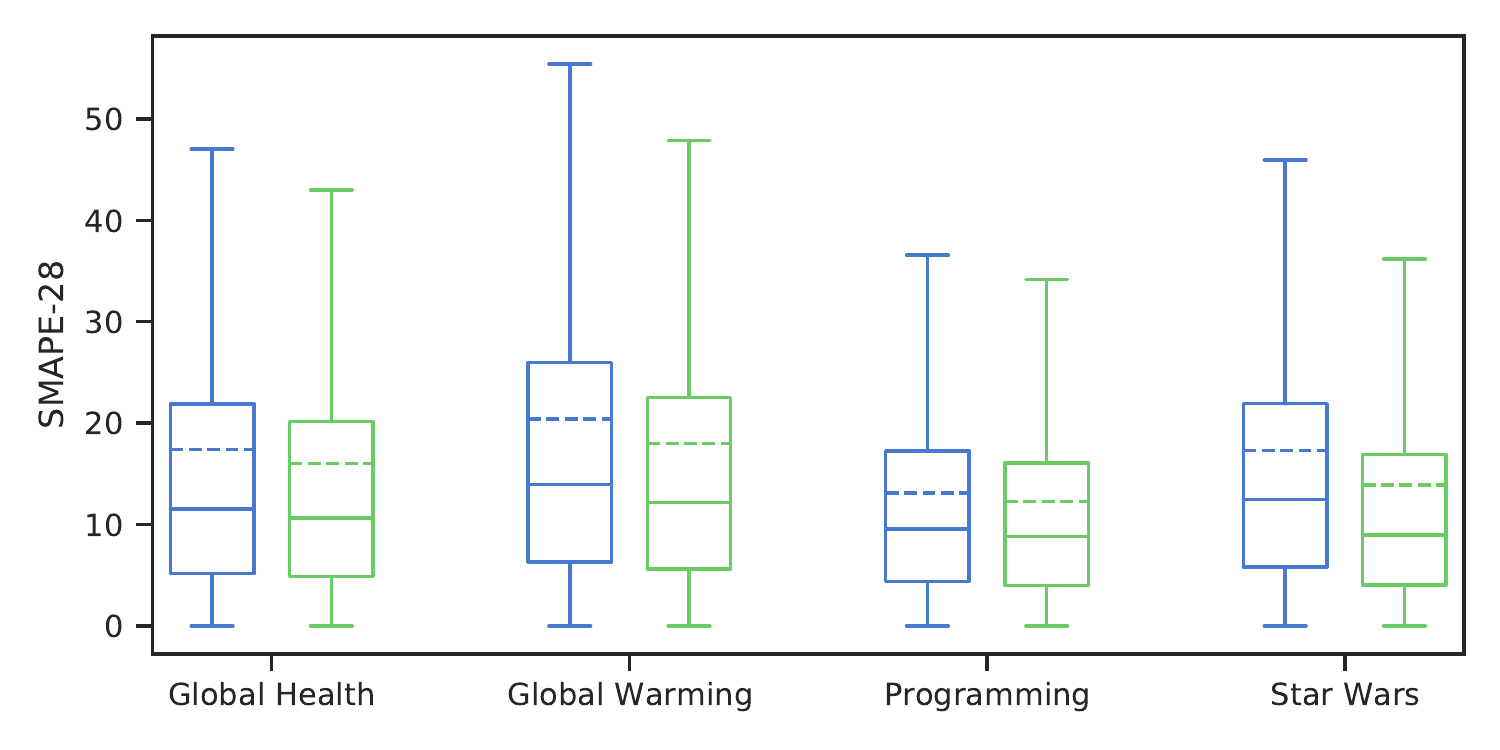}

  \caption{Performance broken down by categories in the test set of \Wiki
  (univariate). The dashed lines indicate mean SMAPE-28. Blue boxplots
  correspond to \modelname-NoNetwork, while green boxplots correspond to
  \modelname\ with one-hop aggregation. All categories benefit from the network
  information.}

  \Description{Performance broken down by page category.}
  \label{fig:subcats}
\end{figure}

On \Wiki, we break our model's performance down by both page category and page
popularity, in order to see if these have any effect on the performance. From
\cref{fig:subcats}, we see that on both \modelname-NoNetwork and \modelname,
there is no significant difference in the performance among the four test
categories---Global Health, Global Warming, Programming, and Star Wars.
Furthermore, the improvement in SMAPE from adding network information is
consistent across these categories.

\begin{figure}[t]
  \centering
  \includegraphics[width=\linewidth]{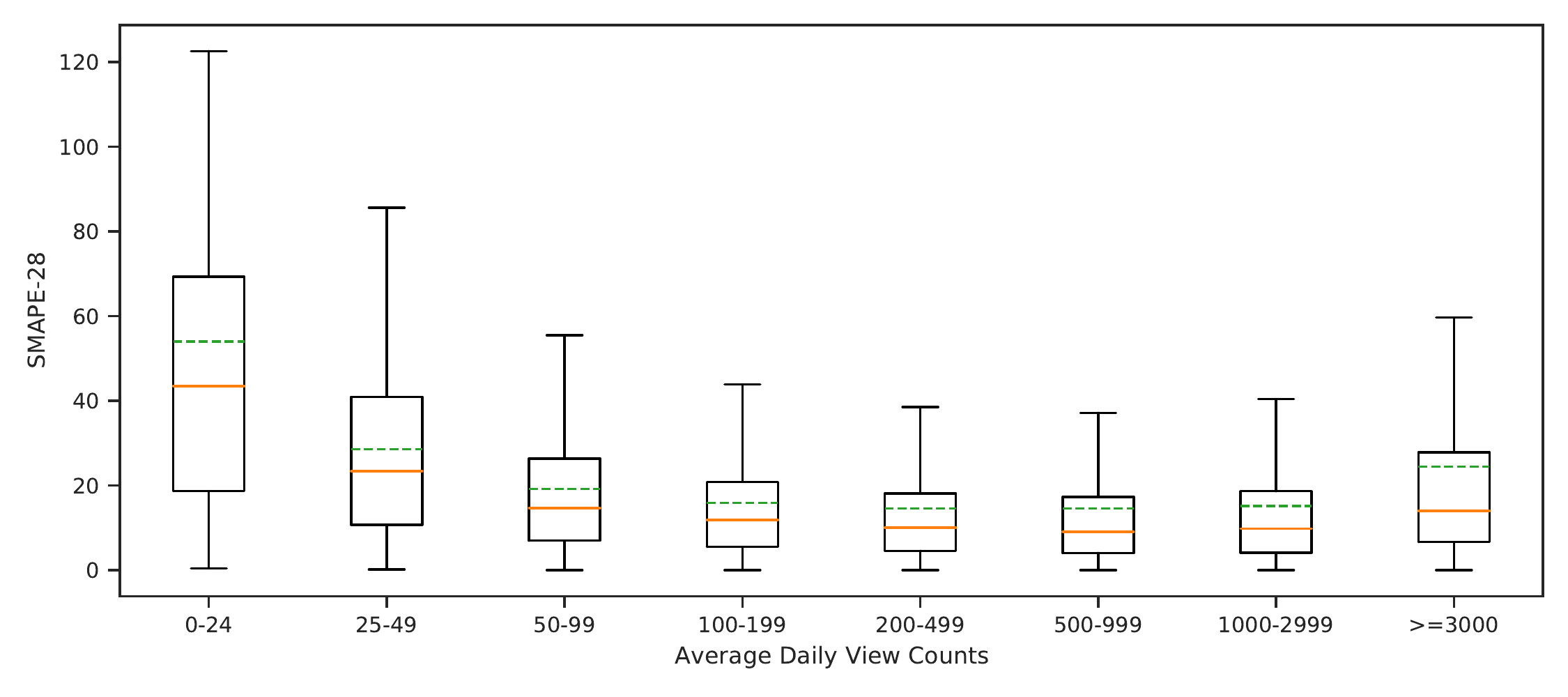}

  \caption{\modelname-NoNetwork's performance on \Wiki (univariate), broken
  down by the popularity of a page. The dashed lines indicate mean SMAPE-28.
  There exists an optimal range of popularity where traffic is the most
  predictable, resulting in a low SMAPE.}

  \Description{Performance broken down by page popularity.}
  \label{fig:traffic_split}
\end{figure}

In contrast, the popularity of a page has a significant impact on how well the
predictions are. From \cref{fig:traffic_split}, we observe that there exists an
optimal range of popularity where traffic is most predictable. Our model is
best at forecasting pages that have between 200 and 1,000 daily visits. Pages
with fewer than 50 daily visits are fairly difficult to forecast. This matches
the observation we made on \SZ, where it is also difficult to forecast the low
traffic speeds. More interestingly, our model's performance also drops slightly
with very popular pages (those with more than 1,000 daily views). This could be
due the traffic of these popular pages being driven mostly by events external
to the network.

\subsection{Hyperparameters and Significance Tests}

This final section provides further details on the hyperparameters and
statistical significance tests. In the main paper, \cref{tab:radflow_params}
shows the hyperparameters of our two key models, \modelname-NoNetwork and
\modelname. \cref{tab:params_ts,tab:params_network} provide hyperparameters of
the remaining model variants. Finally \cref{tab:pvalues1} contains the p-values
from the dependent t-test for paired samples on \Vevo and \Wiki. We note that
if two models differ by more than one digit in the third significant figure, it
is sufficient to conclude that the difference is statistically significant
in these datasets (i.e. p-value < 0.05).

\onecolumn

\begin{table}[htb]
  \caption{Hyperparameters of pure time series models. Rows are numbered
           according to Sec~\ref{ssec:variants}. All model variants use 8
           layers with a dropout of 0.1. For each dataset, we calibrate the
           hidden size so that the number of parameters between all models are
           within 5\% of each other. Note that the N-BEATS model for \Wiki
           (bivariate) consists of two separate models, one predicting the
           desktop traffic and the other predicting the combined mobile/app
           traffic. Each of these two models contains 829,760 parameters.}
  \Description{Hyperparameters of pure time series models.}
  \label{tab:params_ts}
  \begin{tabular}{llrr}
    \toprule
    Dataset & Model & Hidden Size & Parameters \\
    \midrule

    \multirow{3}{*}{\Vevo}
    & (6) LSTM & 164 & 1,625,078 \\
    & (7) N-BEATS & 192 & 1,630,480 \\
    & (8) \modelname-NoNetwork & 128 & 1,589,762 \\
    \midrule

    \multirow{3}{*}{\Wiki (univariate)}
    & (6) LSTM & 164 &  1,625,078  \\
    & (7) N-BEATS & 176 & 1,610,176  \\
    & (8) \modelname-NoNetwork & 128 & 1,589,762  \\
    \midrule

    \multirow{3}{*}{\Wiki (bivariate)}
    & (6) LSTM & 164 & 1,625,900  \\
    & (7) N-BEATS & 120 & 1,659,520  \\
    & (8) \modelname-NoNetwork & 128 & 1,590,020 \\
    \midrule

    \multirow{2}{*}{\LA}
    & (8) \modelname-NoNetwork & 73 & 260,758  \\
    & (9) T-GCN~\cite{Zhao2019TGCNAT} & 292 & 261,060 \\
    \midrule

    \multirow{2}{*}{\SZ}
    & (8) \modelname-NoNetwork & 73 & 260,758 \\
    & (9) T-GCN~\cite{Zhao2019TGCNAT} & 294 & 262,252\\

    \bottomrule
  \end{tabular}
\end{table}

\begin{table}[htb]
  \caption{Hyperparameters of time series models with network information. Rows
  are numbered according to Sec~\ref{ssec:variants}. All model variants use 8
  layers with a dropout of 0.1. For each dataset, we calibrate the hidden size
  so that the number of parameters between all models are within 5\% of each
  other. Some model variants were not evaluated under the two-hop setting
  and are indicated with a hyphen.}
  \Description{Hyperparameters of time series models with network information.}
  \label{tab:params_network}
  \begin{tabular}{llrrrr}
    \toprule
    \multirow{2}{*}{Dataset} & \multirow{2}{*}{Model}
     & \multicolumn{2}{c}{One Hop} & \multicolumn{2}{c}{Two Hops} \\
     \cmidrule(lr){3-4} \cmidrule(lr){5-6}
     & & Hidden Size & Parameters & Hidden Size & Parameters \\
    \midrule

    \multirow{12}{*}{\parbox{3.5cm}{\Vevo \\ \Wiki (univariate)}}
    & (11) LSTM-MeanPooling & 160 & 1,598,562 & 160 & 1,650,082  \\
    & (12) \modelmean & 120 & 1,632,604 & 120 & 1,632,604 \\
    & (13) \modelsage & 118 & 1,606,928 & 118 & 1,634,894 \\
    & (14) \modelgat  & 120 & 1,647,364 & 120 & 1,662,124 \\
    & (15) \modelname & 116 & 1,608,112 & 112 & 1,576,180 \\
    & (16) \modelname ($\bh$ as embeddings) & 124 & 1,586,088 & - & - \\
    & (17) \modelname ($\bp$ as embeddings) & 124 & 1,586,088 & - & - \\
    & (18) \modelname ($\bq$ as embeddings) & 124 & 1,586,088 & - & - \\
    & (19) \modelname ($[\bh; \bp]$ as embeddings) & 116 & 1,632,588 & - & - \\
    & (20) \modelname ($[\bh; \bp; \bq]$  & 104 & 1,639,564 & - & - \\
    & (21) \modelname (no final projection) & 116 & 1,608,112 & - & - \\
    & (22) \modelname (one head) & 116 & 1,608,112 & - & - \\
    \midrule

    \multirow{5}{*}{\Wiki (bivariate)}
    & (11) LSTM-MeanPooling & 160 & 1,599,364 & 160 & 1,650,884 \\
    & (12) \modelmean & 120 & 1,632,968 & 120 &  1,632,968 \\
    & (13) \modelsage & 118 & 1,607,286 & 118 & 1,635,252 \\
    & (14) \modelgat & 120 & 1,647,728 & 120 & 1,662,488 \\
    & (15) \modelname & 116 & 1,608,464 & 112 & 1,576,520 \\
    \midrule

    \LA
    & (15) \modelname & 64 & 260,036 & - & - \\
    \midrule

    \SZ
    & (15) \modelname & 64 & 260,036 & - & - \\

    \bottomrule
  \end{tabular}
\end{table}


\begin{table}[p]
  \caption{Selected p-values from the dependent t-test for paired samples on
  \Vevo and \Wiki. Models are numbered according to Sec~\ref{ssec:variants}. We
  only show pairs where the p-value is at least 0.001. We observe that if two
  models differ by more than one digit in the third significant figure, their
  difference is statistically significant in these datasets ($p < 0.05$).
  }
  \Description{Selected p-values from the dependent t-test for paired samples on
  VevoMusic and WikiTraffic.}
  \label{tab:pvalues1}
  \centerline{
  \begin{tabular}{llllrllrr}
    \toprule
    & & \multicolumn{3}{c}{Group 1} &  \multicolumn{3}{c}{Group 2} & \multirow{2}{*}{p-value} \\
    \cmidrule(lr){3-5} \cmidrule(lr){6-8}
    & & Model & Hops & SMAPE & Model & Hops  & SMAPE & \\
    \midrule
    \multicolumn{2}{c}{\Vevo}
    & (6) LSTM & 0H & 8.68 & (7) N-BEATS & 0H & 8.64 & 0.006 \\
    \cmidrule(lr){1-9}
    \multirow{9}{*}{\rotatebox[origin=c]{90}{\Vevo (static)}}
    & \multirow{5}{*}{\textsc{Forecast}}
    & (12) \modelmean & 1H & 8.34 & (15) \modelname & 1H & 8.33 & 0.171 \\
    & & (13) \modelsage & 1H & 8.39 & (13) \modelsage & 2H & 8.37 & 0.036 \\
    & & (13) \modelsage & 1H & 8.39 & (15) \modelname & 2H & 8.39 & 0.938 \\
    & & (14) \modelgat & 1H & 8.52 & (14) \modelgat & 2H & 8.50 & 0.018 \\
    & & (13) \modelsage & 2H & 8.37 & (15) \modelname & 2H & 8.39 & 0.014 \\
    \cmidrule(lr){2-9}
    & \multirow{4}{*}{\textsc{Imputation}}
    & (11) LSTM-MeanPooling & 1H & 8.14 & (11) LSTM-MeanPooling & 2H & 8.13 & 0.117 \\
    & & (12) \modelmean & 1H & 7.82 & (12) \modelmean & 2H & 7.81 & 0.376 \\
    & & (15) \modelname & 1H & 7.67 & (13) \modelsage & 2H & 7.64 & 0.001 \\
    & & (13) \modelsage & 2H & 7.64 & (15) \modelname & 2H & 7.63 & 0.017 \\
    \midrule
    \multirow{10}{*}{\rotatebox[origin=c]{90}{\Vevo (dynamic)}}
    & \multirow{5}{*}{\textsc{Forecast}}
    & (12) \modelmean & 1H & 8.42 & (13) \modelsage & 1H & 8.43 & 0.618 \\
    & & (12) \modelmean & 1H & 8.42 & (14) \modelgat & 1H & 8.43 & 0.586 \\
    & & (13) \modelsage & 1H & 8.43 & (14) \modelgat & 1H & 8.43 & 0.987 \\
    & & (15) \modelname & 1H & 8.37 & (14) \modelgat & 2H & 8.39 & 0.001 \\
    & & (13) \modelsage & 2H & 8.46 & (15) \modelname & 2H & 8.45 & 0.064 \\
    \cmidrule(lr){2-9}
    & \multirow{5}{*}{\textsc{Imputation}}
    & (11) LSTM-MeanPooling & 1H & 7.91 & (11) LSTM-MeanPooling & 2H & 7.90 & 0.702 \\
    & & (13) \modelsage & 1H & 7.46 & (14) \modelgat & 1H & 7.44 & 0.007 \\
    & & (13) \modelsage & 2H & 7.27 & (14) \modelgat & 2H & 7.28 & 0.091 \\
    & & (13) \modelsage & 2H & 7.27 & (15) \modelname & 2H & 7.27 & 0.871 \\
    & & (14) \modelgat & 2H & 7.28 & (15) \modelname & 2H & 7.27 & 0.110 \\
    \midrule
    \multirow{18}{*}{\rotatebox[origin=c]{90}{\Wiki (univariate)}}
    & \multirow{11}{*}{\textsc{Forecast}}
    & (6) LSTM & 0H & 16.6 & (7) N-BEATS & 0H & 16.6 & 0.471 \\
    & & (6) LSTM & 0H & 16.6 & (11) LSTM-MeanPooling & 2H & 16.7 & 0.002 \\
    & & (6) LSTM & 0H & 16.6 & (13) \modelname-GraphSage & 2H & 16.7 & 0.067 \\
    & & (7) N-BEATS & 0H & 16.6 & (12) \modelname-MeanPooling & 2H & 16.5 & 0.002 \\
    & & (7) N-BEATS & 0H & 16.6 & (13) \modelname-GraphSage & 2H & 16.7 & 0.016 \\
    & & (8) \modelname-NoNetwork & 0H & 16.1 & (14) \modelname-GAT & 1H & 16.2 & 0.019 \\
    & & (8) \modelname-NoNetwork & 0H & 16.1 & (15) \modelname & 1H & 16.2 & 0.027 \\
    & & (12) \modelmean & 1H & 16.5 & (12) \modelmean & 2H & 16.5 & 0.105 \\
    & & (14) \modelgat & 1H & 16.2 & (15) \modelname & 1H & 16.2 & 0.948 \\
    & & (11) LSTM-MeanPooling & 2H & 16.7 & (13) \modelsage & 2H & 16.7 & 0.139 \\
    & & (14) \modelgat & 2H & 16.0 & (15) \modelname & 2H & 16.0 & 0.934 \\
    \cmidrule(lr){2-9}
    & \multirow{7}{*}{\textsc{Imputation}}
    & (12) \modelmean & 1H & 15.1 & (14) \modelgat & 1H & 15.1 & 0.844 \\
    & & (12) \modelmean & 1H & 15.1 & (13) \modelsage & 2H & 15.0 & 0.001 \\
    & & (14) \modelgat & 1H & 15.1 & (12) \modelmean & 2H & 15.1 & 0.001 \\
    & & (14) \modelgat & 1H & 15.1 & (13) \modelsage & 2H & 15.0 & 0.001 \\
    & & (11) LSTM-MeanPooling & 2H & 15.2 & (12) \modelmean & 2H & 15.1 & 0.098 \\
    & & (11) LSTM-MeanPooling & 2H & 15.2 & (14) \modelgat & 2H & 15.2 & 0.625 \\
    & & (12) \modelmean & 2H & 15.1 & (14) \modelgat & 2H & 15.2 & 0.064 \\
    \midrule
    \multirow{12}{*}{\rotatebox[origin=c]{90}{\Wiki (bivariate)}}
    & \multirow{8}{*}{\textsc{Forecast}}
    & (7) N-BEATS & 0H & 20.3 & (11) LSTM-MeanPooling & 1H & 20.2 & 0.013 \\
    & & (7) N-BEATS & 0H & 20.3 & (12) \modelname-MeanPooling & 2H & 20.2 & 0.015 \\
    & & (8) \modelname-NoNetwork & 0H & 19.4 & (13) \modelname-GraphSage & 1H & 19.4 & 0.112 \\
    & & (11) LSTM-MeanPooling & 1H & 20.2 & (12) \modelmean & 2H & 20.2 & 0.680 \\
    & & (15) \modelname & 1H & 19.9 & (11) LSTM-MeanPooling & 2H & 19.9 & 0.380 \\
    & & (15) \modelname & 1H & 19.9 & (13) \modelsage & 2H & 19.9 & 0.172 \\
    & & (11) LSTM-MeanPooling & 2H & 19.9 & (13) \modelsage & 2H & 19.9 & 0.104 \\
    & & (14) \modelgat & 2H & 19.7 & (15) \modelname & 2H & 19.6 & 0.487 \\
    \cmidrule(lr){2-9}
    & \multirow{4}{*}{\textsc{Imputation}}
    & (12) \modelmean & 1H & 18.5 & (15) \modelname & 2H & 18.5 & 0.596 \\
    & & (14) \modelgat & 1H & 18.3 & (15) \modelname & 1H & 18.3 & 0.825 \\
    & & (14) \modelgat & 1H & 18.3 & (13) \modelsage & 2H & 18.4 & 0.025 \\
    & & (15) \modelname & 1H & 18.3 & (13) \modelsage & 2H & 18.4 & 0.019 \\
    \bottomrule
  \end{tabular}}
\end{table}